\documentclass[fleqn,10pt]{article}
\usepackage{fullpage}
\usepackage{amsfonts}
\usepackage{amsmath}
\usepackage{multirow}
\usepackage{longtable}
\usepackage{graphicx}
\usepackage{sidecap}
\usepackage{subfigure}
\usepackage[affil-it]{authblk}

\def\kmin{k_{\mathrm{min}}}

\def\pn{p_n}
\def\pr{p_r}
\def\kvec{\mathbf{k}}
\def\Pk{P({\kvec})}
\def\Fkm{F_{\kvec, m}}
\def\skm{s_{\kvec, m}}
\def\skmo{s_{\kvec, m-1}}
\def\dskm{\dot{s}_{\kvec, m}}
\def\sumk{\sum_{\kvec}}
\def\summ{\sum_m}
\def\bs{\beta_s}
\def\Bkm{B_{k, m}}
\def\Bkom{B_{k-1, m}}
\def\Bkmo{B_{k, m-1}}
\def\ft{f_t}

\newcommand{\be}{\begin{equation}}
\newcommand{\ee}{\end{equation}}

\def\pn{p_n}
\def\pr{p_r}
\def\kvec{\mathbf{k}}
\def\Pk{P({\kvec})}
\def\Fkm{F_{\kvec, m}}
\def\skm{s_{\kvec, m}}
\def\skmo{s_{\kvec, m-1}}
\def\dskm{\dot{s}_{\kvec, m}}
\def\pk{\rho_{\kvec}}
\def\sumk{\sum_{\kvec}}
\def\summ{\sum_m}
\def\sumkc{\sum_{\kvec | c \neq 0}}
\def\summLess{\sum_{m < k\phi_c}}
\def\summMore{\sum_{m \geq k\phi_c}}
\def\bs{\beta_s}
\def\Bkm{B_{k, m}}
\def\Bkom{B_{k-1, m}}
\def\Bkmo{B_{k, m-1}}
\def\ft{f_t}

\begin{document}

\title{\textbf{Local cascades induced global contagion: \\ How heterogeneous thresholds, exogenous effects, and unconcerned behaviour govern online adoption spreading}}

\author[1]{M\'arton Karsai
\thanks{Corresponding author: marton.karsai@ens-lyon.fr}}
\affil[1]{\normalsize{Laboratoire de l'Informatique du Parall\'elisme, INRIA-UMR 5668, IXXI, ENS de Lyon, 69364 Lyon, France}}

\author[2,3]{Gerardo I\~niguez}
\affil[2]{\normalsize{Department of Computer Science, School of Science, Aalto University, 00076, Finland}}
\affil[3]{\normalsize{Centro de Investigaci{\'o}n y Docencia Econ{\'o}micas, CONACYT, 01210 M{\'e}xico D.F., Mexico}}

\author[4,5]{Riivo Kikas}
\affil[4]{\normalsize{Institute of Computer Science, University of Tartu, 50409 Tartu, Estonia}}
\affil[5]{\normalsize{Software Technology and Applications Competence Center (STACC), 51003 Tartu, Estonia}}

\author[2]{Kimmo Kaski}

\author[6,7]{J\'anos Kert\'esz}
\affil[6]{\normalsize{Center for Network Science, Central European University, 1051 Budapest, Hungary}}
\affil[7]{\normalsize{Institute of Physics, Budapest University of Technology and Economics, 1111 Budapest, Hungary}}

\date{}

\maketitle

\textbf{keywords:} cascading behaviour, social spreading phenomena, complex contagion, adoption thresholds

\begin{abstract}
Adoption of innovations, products or online services is commonly interpreted as a spreading process driven to large extent by social influence and conditioned by the needs and capacities of individuals. To model this process one usually introduces behavioural threshold mechanisms, which can give rise to the evolution of global cascades if the system satisfies a set of conditions. However, these models do not address temporal aspects of the emerging cascades, which in real systems may evolve through various pathways ranging from slow to rapid patterns. Here we fill this gap through the analysis and modelling of product adoption in the world's largest voice over internet service, the social network of Skype. We provide empirical evidence about the heterogeneous distribution of fractional behavioural thresholds, which appears to be independent of the degree of adopting egos. We show that the structure of real-world adoption clusters is radically different from previous theoretical expectations, since vulnerable adoptions---induced by a single adopting neighbour---appear to be important only locally, while spontaneous adopters arriving at a constant rate and the involvement of unconcerned individuals govern the global emergence of social spreading.
\end{abstract}

\section*{Introduction}

Spreading of opinions, frauds, behavioural patterns, and product adoptions are all examples of social contagion phenomena where collective patterns emerge due to correlated decisions of a large number of individuals. Although these choices are personal, they are not independent but potentially driven by several processes such as social influence \cite{Centola2010Spread}, homophily \cite{McPherson2001}, and information arriving from external sources like news or mass media \cite{Toole2011Modeling}. Social contagion evolves over networks of interconnected individuals, where links associated with social ties transfer influence between peers \cite{RevModPhys.81.591}. Several earlier studies aimed to identify the dominant mechanisms at play in social contagion processes \cite{Rogers2003Diffusion, Granovetter1978Threshold,schelling1969models,Axelrod1997Dissemination}. One key element, termed behavioural threshold by Granovetter \cite{Granovetter1978Threshold}, is defined as \textit{``the number or proportion of others who must make one decision before a given actor does so''}. Following this idea various network models have been introduced \cite{Watts2002Simple,Handjani1997Survival,valente-thresholds-1996,Watts2007Influentials,Melnik2013Multistage,Gomez2010Modeling} to understand the threshold-driven spreading, commonly known as \textit{complex contagion} \cite{Centola2007Complex}. Although these models are related to a larger set of collective dynamics, they are particularly different from \textit{simple contagion} where the exposure of nodes is driven by independent contagion stimuli \cite{Barrat2008Dynamical, Bass1969}. In addition, collective adoption patterns may appear as a consequence of homophilic structural correlations, where connected individuals adopt due to their similar interests and not due to direct social influence. Distinguishing between the effects of social influence and homophily at the individual level remains as a challenge \cite{Aral2009,Shalizi2011}. Furthermore, in real social spreading phenomena all these mechanisms are arguably present. However, while in the case of homophily the  
adoption behaviour is only seemingly correlated, and for simple contagion only the number of exposures matters, in complex contagion the fraction of adopting neighbours relative to the total number of partners determines whether a node adopts or not, capturing the natural mechanisms involved in individuals' decision makings \cite{Holt06,bikhchandani-hirshleifer-welch-92, Karsai2014Complex}. Due to this additional complexity, threshold models are able to emulate system-wide adoption patterns known as global cascades.

Behavioural cascades are rare but potentially stupendous social spreading phenomena, where collective patterns of exposure emerge as a consequence of small initial perturbations. Some examples are the rapid emergence of political and grass-root movements \cite{GonzalezBailon2011Dynamics,BorgeHolthoefer2011Structural,EllisInformation}, fast spreading of information \cite{Dow2013Anatomy,Gruhl2004Information,Banos2013Role, Watts2007Influentials,Hale2013Regime,Leskovec2005Patterns,Leskovec2007Dynamics,Goel2012Structure} or behavioural patterns \cite{Fowler2009Cooperative}, etc. The characterisation \cite{Goel2012Structure, BorgeHolthoefer2013Cascading,Hackett2013Cascades,Gleeson2008Cascades,Brummitt2011,GhoshCascadesArxiv2010} and modelling \cite{Watts2002Simple,Hurd2013Watts, Singh2013Thresholdlimited,Gleeson2007Seed} of such processes have received plenty of attention and provide some basic understanding of the conditions and structure of empirical and synthetic cascades. However, these studies commonly fail in addressing the temporal dynamics of the emerging cascades, which may vary considerably between different cases of social contagion. Moreover, they have not answered why real-world cascades can evolve through various dynamic pathways ranging from slow to rapid patterning, especially in systems where the threshold mechanisms play a role and social phenomena spread globally. Besides the case of rapid cascading mentioned above, an example of the other extreme is the propagation of products in social networks \cite{Bass1969}, where adoption evolves gradually even if it is driven by threshold mechanisms and may cover a large fraction of the total population \cite{Karsai2014Complex}. This behaviour characterises the adoption of online services such as Facebook, Twitter, LinkedIn and Skype (Fig.\ref{fig:1}a), since their yearly maximum relative growth of cumulative adoption \cite{SocialMedia} (for definition see Material and Methods (MM)) is lower than in the case of rapid cascades as suggested e.g. by the Watts threshold (WT) model.

To fill this gap in the modelling of social diffusion, here we will analyse and model real-world examples of social contagion phenomena. Our aim is to identify the crucial mechanisms necessary to consider in models of complex contagion to match them better with reality, and define a model that incorporates these mechanisms and captures the possible dynamics leading to the emergence of real-world global cascades. We follow the adoption dynamics of the Skype paid service ``buy credit'' for $89$ months since 2004, which evolves over the social network of one of the largest voice over internet providers in the world. Data includes the time of first payment of each user, an individual and conscious action that tracks adoption behaviour. In addition we follow the ``subscription" service over $42$ months since 2008 (for results see Supplementary Information [SI]). In contrast to other empirical studies where incomplete knowledge about the underlying social network leads to unavoidable bias \cite{Karsai2014Complex},  we use here the largest connected component of the aggregated free Skype service as the underlying structure, where nodes are Skype users and links confirmed contacts between them. This is a good approximation since it maps all connections in the Skype social network without sampling, and the paid service is only available for individuals already enrolled in the Skype network. The underlying structure is an aggregate from September 2003 to November 2011 (i.e. over $99$ months) and contains roughly 4.4 billion links and 510 million registered users worldwide \cite{SkypeIPO}. The data is fully anonymised and considers only confirmed connections between users (for more data details see SI).

In what follows we first provide empirical evidence of the distribution of individual adoption thresholds and other structural and dynamical features of a worldwide adoption cluster. We incorporate the observed structural and threshold heterogeneities into a dynamical threshold model where multiple nodes adopting spontaneously (i.e. firstly among their neighbours) are allowed \cite{Ruan2015}. We find that if the fraction of users who reject to adopt the product is large, the system enters a quenched state where the evolution and structure of the global adoption cluster is very similar to our empirical observations. Model calculations and the analysis of the real social contagion process suggest that the evolving structure of an adoption cluster differs radically from what has been proposed earlier \cite{Watts2002Simple}, since it is triggered by several spontaneous adoptions arriving at a constant rate, while stable adopters who are initially resisting exposure, are actually responsible for the emergence of global social adoption (Fig.~\ref{fig:1}b and c).

\section*{Results}

Social contagion phenomena can be modelled as binary-state processes evolving on networks and driven by threshold mechanisms. In these systems individuals are represented by nodes, each being either in a susceptible (0) or adopter (1) state and influencing each other by transferring information via social ties \cite{Granovetter1978Threshold}. Nodes are connected in a network with degree distribution $P(k)$ and average degree $z = \langle k \rangle$. In addition, each node has an individual threshold $\phi \in [0, 1]$ drawn from a distribution $P(\phi)$ with average $w = \langle \phi \rangle$. This threshold determines the minimum fraction of exposed neighbours that triggers adoption and captures the resistance of an individual against engaging in spreading behaviour. Once a node reaches its threshold, it switches state from $0$ to $1$ and keeps it until the end of the dynamics. In his seminal paper about threshold dynamics, Watts \cite{Watts2002Simple} classified nodes into three categories based on their threshold and degree. He identified {\it innovator} nodes that spontaneously change state to $1$, thus starting the process. Such nodes have a trivial threshold $\phi=0$. Then there are nodes with threshold $0 < \phi \leq 1/k$, called {\it vulnerable}, which need one adopting neighbour before their own adoption. Finally, there are more resilient nodes with threshold $\phi>1/k$, denoted as {\it stable}, referring to individuals in need of strong social influence to follow the actions of their acquaintances.

In the WT model \cite{Watts2002Simple}, small perturbations (like the spontaneous adoption of a single seed node) can trigger global cascading patterns.  However, their emergence is subject to the so-called {\it cascade condition}: the innovator seed has to be linked to a percolating vulnerable cluster, which adopts immediately afterwards and further triggers a global cascade (i.e. a set of adopters larger than a fixed fraction of the finite network). The cascade condition is satisfied if the network is inside a bounded regime in $(w, z)$-space \cite{Watts2002Simple}. This regime depends on degree and threshold heterogeneities \cite{Watts2002Simple} and may change its shape if several innovators start the process \cite{Singh2013Thresholdlimited}.

\subsection*{Empirical observations}

\begin{figure*}[t]
\centering
\includegraphics[width=0.9\textwidth,angle=0]{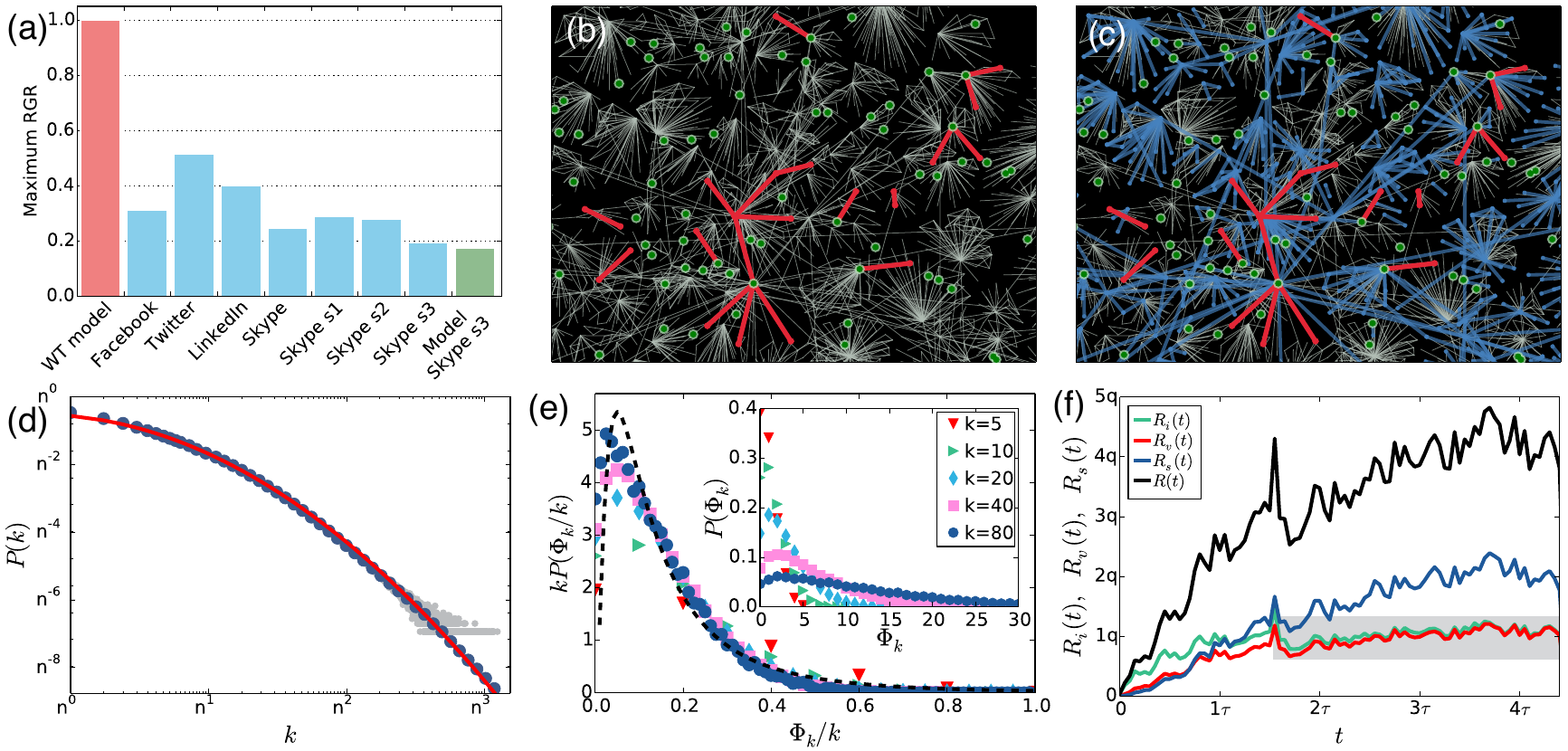}
\caption{\textbf{Structure and dynamics of online service adoption.} \textbf{(a)} Yearly maximum relative growth rate (RGR) of cumulative adoption (see MM) for several online social-communication services \cite{SocialMedia}, including three Skype paid services (s1 - "subscription", s2  - "voicemail", and s3 - "buy credit"). The red bar corresponds to a rapid cascade of adoption suggested by the Watts threshold (WT) model, while the green bar is the model prediction for Skype s3. \textbf{(b-c)} Snowball sample of the Skype social network (gray links) with nodes and links coloured according to their adoption state: multiple innovators (green nodes), induced small vulnerable trees (red nodes and links), and the triggered connected stable cluster (blue nodes and links). Note that some vulnerable and stable clusters seemingly appear without an innovator seed due to the finite distance used in the snowball sampling method. \textbf{(d)} Degree distribution $P(k)$ of the Skype network (gray/blue circles for raw/binned data) on double log-scale with arbitrary base $n$. $P(k)$ is fitted by a lognormal distribution (see MM and SI) with parameters $\mu_D=1.2$ and $\sigma_D=1.39$, and average $z = 8.56$ (red line). \textbf{(e)} Distribution $P(\Phi_k)$ of integer thresholds $\Phi_k$ for several degree groups in Skype s3 (inset). By using $P(\Phi_k, k) = k P(\Phi_k/k)$, these curves collapse to a master curve approximated by a lognormal function (dashed line in main panel) with parameters $\mu_T=-2$ and $\sigma_T=1$, as constrained by the average threshold $w = 0.19$ (see MM and SI). \textbf{(f)} Adoption rate of innovators [$R_i(t)$], vulnerable nodes [$R_v(t)$], and stable nodes [$R_s(t)$], as well as net service adoption rate [$R(t)$]. Rates are measured with a 1-month time window, while $q$ and $\tau$ are arbitrary constants. The shaded area indicates the regime where innovators adopt approximately with constant rate.
\label{fig:1}}
\end{figure*}

Degree and threshold heterogeneities are indeed present in the social network of Skype. The degree distribution $P(k)$ is well approximated by a lognormal function $P(k) \propto k^{-1} e^{-(\ln k - \mu_D)^2/(2\sigma_D^2)}$ ($k \geq \kmin$) with parameters $\mu_D=1.2$, $\sigma_D=1.39$ and $\kmin=1$ (Fig.~\ref{fig:1}d), giving an average degree $z = 8.56$ (for goodness of fit see SI). Moreover, at the time of adoption we can measure the threshold $\phi=\Phi_k/k$ of a user by counting the number $\Phi_k$ of its neighbours who have adopted the service earlier. We then group users by degree and calculate the distribution $P(\Phi_k)$ of the integer threshold $\Phi_k$ \cite{Gleeson2008Cascades} (Fig.~\ref{fig:1}e). By using the scaling relation $P(\Phi_k, k) = k P(\Phi_k/k)$ all distributions collapse to a master curve well approximated by a lognormal function $P(\phi) \propto \phi^{-1} e^{-(\ln\phi - \mu_T)^2/(2\sigma_T^2)}$, with parameters $\mu_T=-2$ and $\sigma_T=1$ as constrained by the average threshold $w = 0.19$ (see MM and SI). Note that we observe qualitatively the same scaling and lognormal shape of the threshold distribution for another service (see SI). These empirical observations, in addition to the broad degree distribution, provide quantitative evidence about the heterogeneous nature of adoption thresholds.

Since we know the complete structure of the online social network, as well as the first time of service usage for all adopters, we can follow the temporal evolution of the adoption dynamics. By counting the number of adopting neighbours of an ego, we identify innovators ($\Phi_k=0$), and vulnerable ($\Phi_k=1$) or stable ($\Phi_k>1$) nodes. The adoption rates for these categories behave rather differently from previous suggestions \cite{Watts2002Simple} (Fig.~\ref{fig:1}f). First, there is not only one seed but an increasing fraction of innovators in the system who, after an initial period, adopt approximately at a constant rate. Second, vulnerable nodes adopt approximately with the same rate as innovators suggesting a strong correlation between these types of adoption. Third, the overall adoption process accelerates due to the increasing rate of stable adoptions induced by social influence. At the same time a giant adoption cluster grows and percolates through the whole network (Fig.~\ref{fig:3}a, main panel). Despite of this expansion dynamics and connected structure of the service adoption cluster, the service reaches less than $6\%$ of the total number of active Skype users over a period of $7$ years \cite{SkypeIPO}. Therefore we ask whether one can refer to these adoption clusters as cascades. They are not triggered by a small perturbation but induced by several innovators; their evolution is not instantaneous but ranges through several years; and although they involve millions of individuals, they reach only a reduced fraction of the whole network. To answer we incorporate the above mentioned features into a dynamical threshold model \cite{Ruan2015} with a growing group of innovators and investigate their effect on the evolution of global social adoption. Note that we also perform a null model study to demonstrate, on the system level, that social influence dominates the contagion process, but not homophily (see section S3 of the SI, together with another empirical spreading scenario in S7.1).

\subsection*{Model}

Our modelling framework is an extension to conventional threshold dynamics on networks studied by Watts, Gleeson, Singh, and others, where all nodes are initially susceptible and innovators are only introduced as an initial seed of arbitrary size~\cite{Watts2002Simple,Gleeson2007Seed,Singh2013Thresholdlimited}. Apart from the threshold rule discussed above, our model considers two additional features: (i) a fraction $r$ of `immune' nodes that never adopt, indicating lack of interest in the service; (ii) due to external influence, susceptible nodes adopt the innovation spontaneously (i.e. become innovators) throughout time with constant rate $p_n$, rather than only at the beginning of the dynamics. In this way, the dynamical evolution of the system is completely defined by the online social network, the distribution $P(\phi)$ and the parameters $r$, $p_n$. For the sake of simplicity we consider a configuration-model network, i.e., we ignore correlations in the social network and characterise it solely by its degree distribution $P(k)$. Furthermore, node degrees and thresholds are considered to be independent \cite{Gleeson2008Cascades,gleeson2013binary,gleeson2011high}.

Our threshold model, which has also been introduced in \cite{Ruan2015}, can be studied analytically by extending the framework of approximate master equations (AMEs) for monotone binary-state dynamics recently developed by Gleeson~\cite{Gleeson2008Cascades,gleeson2013binary,gleeson2011high}, where the transition rate between susceptible and adoption states only depends on the number $m$ of network neighbours that have already adopted. We describe a node by the property vector $\kvec = (k, c)$, where $k = k_0, k_1, \ldots k_{M-1}$ is its degree and $c = 0, 1, \ldots, M$ its type, i.e. $c = 0$ is the type of the fraction $r$ of immune nodes, while $c \neq 0$ is the type of all non-immune nodes that have threshold $\phi_c$. In this way $P(\phi)$ is substituted by the discrete distribution of types $P(c)$ (for $c > 0$). The integer $M$ is the maximum number of degrees (or non-zero types) considered in the AME framework, which can be increased to improve the accuracy of the analytical approximation at the expense of speed in its numerical computation (see S4.2). Under these conditions, the AME system describing the dynamics of the threshold model is reduced to the pair of ordinary differential equations (see SI),
\begin{subequations}
\label{eq:reducedAMEs}
\begin{align}
\dot{\rho} &= h(\nu, t) - \rho, \\
\dot{\nu} &= g(\nu, t) - \nu,
\end{align}
\end{subequations}
where $\rho(t)$ is the fraction of adopters in the network, $\nu(t)$ is the probability that a randomly chosen neighbour of a susceptible node is an adopter, and the initial conditions are $\rho(0) = \nu(0) = 0$. Here,
\begin{equation}
\label{eq:hTerm}
h = (1 - r) \Big[ \ft + (1 - \ft) \sum_{\kvec | c \neq 0} P(k) P(c) \sum_{m \geq k\phi_c} \Bkm(\nu) \Big],
\end{equation}
and,
\begin{equation}
\label{eq:gTerm}
g = (1 - r) \Big[ \ft + (1 - \ft) \sum_{\kvec | c \neq 0} \frac{k}{z} P(k) P(c) \sum_{m \geq k\phi_c} \Bkom(\nu) \Big],
\end{equation}
where $\ft = 1 - (1 - \pr) e^{-\pr t}$, $\pr = p_n / (1 - r)$, and $\Bkm(\nu) = \binom{k}{m} \nu^m (1 - \nu)^{k - m}$ is the binomial distribution. The fraction of adopters $\rho$ is then obtained by solving Eq.~(\ref{eq:reducedAMEs}) numerically. Since susceptible nodes adopt spontaneously with rate $p_n$, the fraction of innovators $\rho_0(t)$ in the network is given by (see S4.3),
\begin{equation}
\label{eq:innovFrac}
\rho_0 = \pr \int_0^t (1 - r - \rho) dt.
\end{equation}

\begin{figure}[t]
\centering
\includegraphics[width=0.6\textwidth,angle=0]{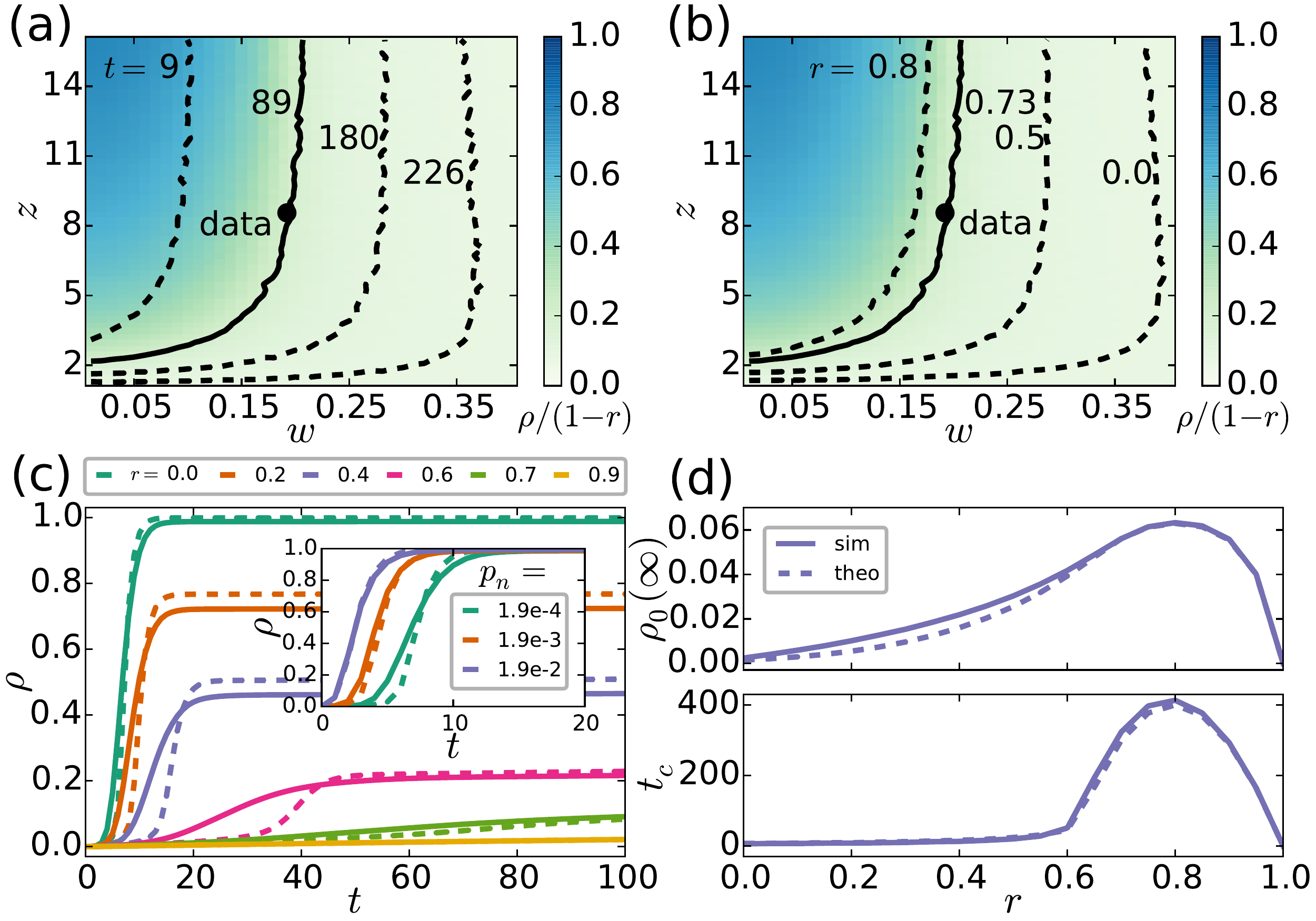}
\caption{{\bf Threshold model for the adoption of online services.} {\bf (a-b)} Surface plot of the normalised fraction of adopters $\rho / (1 - r)$ in $(w, z)$-space, for $r = 0.73$ and $t = 89$. Contour lines signal parameter values for which $20\%$ of non-immune nodes have adopted, for fixed $r$ and varying time (a), and for fixed time and varying $r$ (b). The continuous contour line and dot indicate parameter values in the last observation of Skype s3. A regime of maximal adoption ($\rho \approx 1 - r$) grows as time goes by, and shrinks for larger $r$. {\bf (c)} Time series of the fraction of adopters $\rho$ for fixed $p_n = 0.00019$ and varying $r$ (main), and for fixed $r = 0$ and varying $p_n$ (inset). These curves are well approximated by the solution of Eq.~(\ref{eq:reducedAMEs}) for $k_0 = 3$, $k_{M-1} = 150$ and $M = 25$ (dashed lines). The dynamics is clearly faster for larger $p_n$. As $r$ increases, the system enters a regime where the dynamics is slowed down and adopters are mostly innovators. {\bf (d)} Final fraction of innovators $\rho_0(\infty)$ and time $t_c$ when $50\%$ of non-immune nodes have adopted as a function of $r$, both simulated and theoretical. The crossover to a regime of slow adoption is characterised by a maximal fraction of innovators and time $t_c$. Unless otherwise stated, $p_n=0.00019$ and we use $N=10^4$, $\mu_D=1.09$, $\sigma_D=1.39$, $\kmin=1$, $\mu_T=-2$, and $\sigma_T=1$ to obtain $z = 8.56$ and $w = 0.19$ as in Skype s3. The difference in $\mu_D$ between data and model is due to finite-size effects (see Materials and Methods). Numerical results are averages over $10^2$ (a-b) and $10^3$ (c-d) realisations.
\label{fig:2}}
\end{figure}

We also implement the threshold model numerically via a Monte Carlo simulation in a network of size $N$, with a lognormal degree distribution and a lognormal threshold distribution as observed empirically. Thus, we can explore the behaviour of $\rho$ and $\rho_0$ as a function of $z$, $w$, $p_n$ and $r$, both in the numerical simulation and in the theoretical approximation given by Eqs.~(\ref{eq:reducedAMEs}) and~(\ref{eq:innovFrac}). For $p_n > 0$ some nodes adopt spontaneously as time passes by, leading to a frozen state characterised by a final fraction $\rho(\infty) = 1 - r$ of adopters. However, the time needed to reach such state depends heavily on the distribution of degrees and thresholds, as signalled by a region of large adoption ($\rho \approx 1 - r$) that grows in $(w, z)$-space with time (contour lines in Fig.~\ref{fig:2}a). If we fix a time in the dynamics and vary the fraction of immune nodes instead, this region shrinks as $r$ increases (contour lines in Fig.~\ref{fig:2}b). In other words, the set of networks (defined by their average degree and threshold) that allow the spread of adoption is larger at later times in the dynamics, or when the fraction of immune nodes is small. When both $t$ and $r$ are fixed, the normalised fraction of adopters $\rho / (1 - r)$ gradually decreases for less connected networks with larger thresholds (surface plot in Fig.~\ref{fig:2}a and b).

For $r \approx 0$ the critical fraction of innovators necessary to trigger a cascade of fast adoption throughout all susceptible nodes may be identified as the inflection point in the time series of $\rho$ (Fig.~\ref{fig:2}c, inset). The adoption cascade appears sooner for larger $p_n$, since this parameter regulates how quickly the critical fraction of innovators is reached. Yet as we increase $r$ above a threshold $r_c$, the system enters a regime where rapid cascades disappear and adoption is slowed down. The crossover between these regimes is gradual, as seen in the shape of $\rho$ for increasing $r$ (Fig.~\ref{fig:2}c, main panel). We may identify $r_c$ in various ways: by the maximum in both the final fraction of innovators $\rho_0(\infty)$ and the critical time $t_c$ when $\rho = (1-r)/2$ (Fig.~\ref{fig:2}d), or as the $r$ value where the inflection point in $\rho$ disappears. These measures indicate $r_c \approx 0.8$ for the chosen parameters. All global properties of the dynamics (like the functional dependence of $\rho$ and $\rho_0$) are very well approximated by the solution of Eqs.~(\ref{eq:reducedAMEs}) and~(\ref{eq:innovFrac}) (dashed lines in Fig.~\ref{fig:2}c and d). Indeed, the AME framework is able to capture the shape of the $\rho$ time series, the crossover between regimes of fast and slow adoption, as well as the maximum in $\rho_0(\infty)$ and $t_c$.

\subsection*{Validation} 

\begin{figure}
\centering
\includegraphics[width=0.6\textwidth,angle=0]{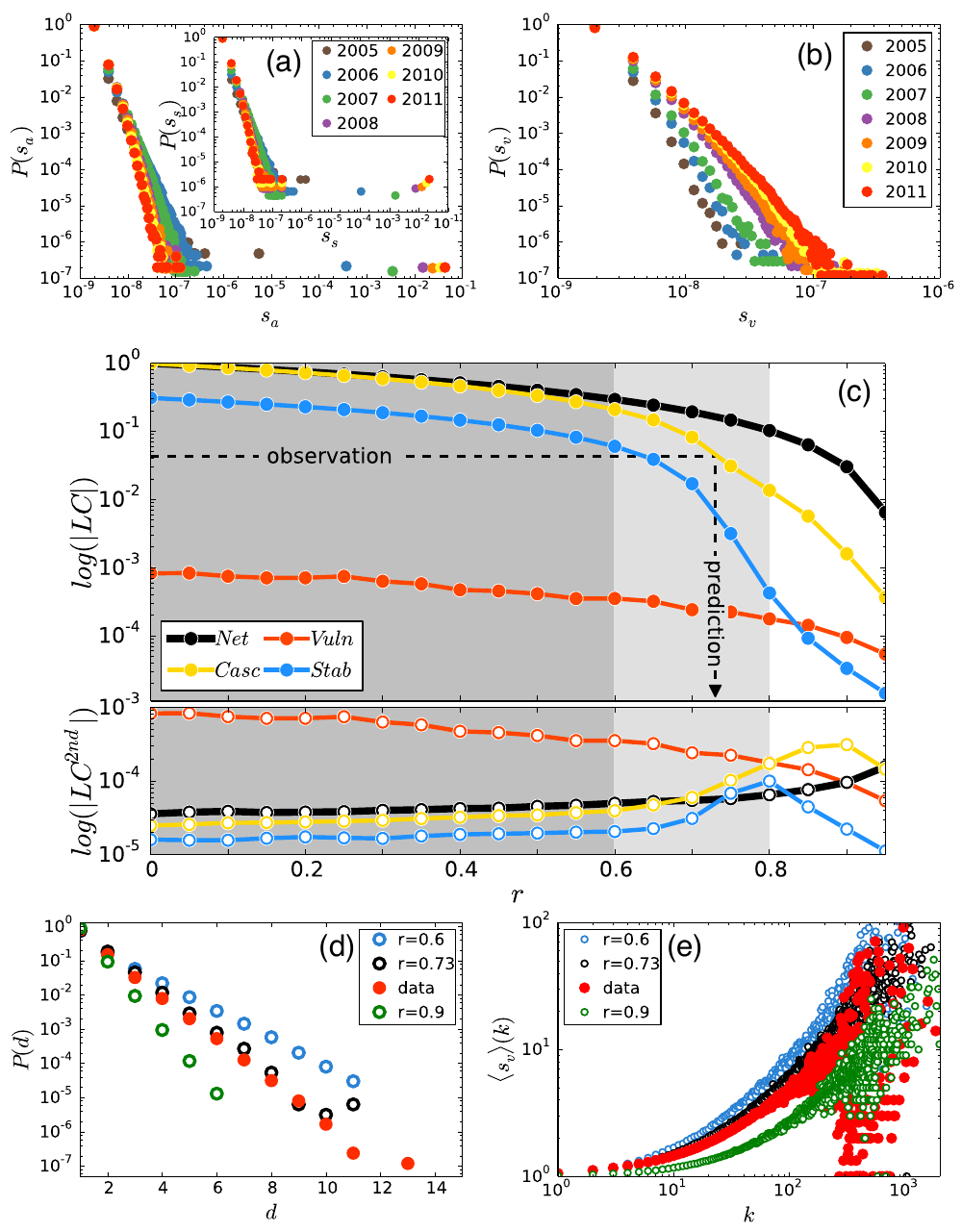}
\caption{\textbf{Empirical cluster statistics and simulation results.} \textbf{(a)} Empirical connected-component size distribution at different times for the adoption [$P(s_a)$, main panel] and stable adoption [$P(s_s)$, inset] networks, with $s_a$ and $s_s$ relative to system size. \textbf{(b)} Empirical connected-component size distribution $P(s_v)$ for the relative size of innovator-induced vulnerable trees at different times. \textbf{(c)} Average size of the largest ($LC$) and 2nd largest ($LC^{2nd}$) components of the model network (`Net'), adoption network (`Casc'), stable network (`Stab'), and induced vulnerable trees (`Vuln') as a function of $r$. Dashed lines show the observed relative size of the real $LC$ of the adopter network in $2011$ [see main panel in (a)] and the predicted $r$ value. \textbf{(d)} Distribution $P(d)$ of depths of induced vulnerable trees in both data and model for several $r$ values, showing a good fit with the data for $r=0.73$. The difference in the tail is due to finite-size effects. \textbf{(e)} Correlation $\langle s_v \rangle (k)$ between innovator degree and average size of vulnerable trees in both data and model with the same $r$ values as in (e). Model calculations for (d) and (e) correspond to networks of size $N=10^6$ and are averaged over $10^2$ realisations.
\label{fig:3}}
\end{figure}

To better understand how innovation spreads throughout real social networks, we take a closer look at the internal structure of the service adoption process. By taking into account individual adoption times we construct an evolving adoption network with links between users who have adopted the service before time $t$ and are connected in the social structure. In order to avoid the effect of instantaneous group adoptions (evidently not driven by social influence), we only consider links between nodes who are neighbours in the underlying social network and whose adoption did not happen at the same time. This way links in the adoption graph indicate ties where social influence among individuals could have existed. The size distribution $P(s_a)$ of connected components in the adoption network shows the emergence of a giant percolating component over time (Fig.~\ref{fig:3}a), along with several other small clusters. Moreover, after decomposition we observe that the giant cluster does not consist of a single innovator seed and percolating vulnerable tree~\cite{Watts2002Simple}, but builds up from several innovator seeds that induce small vulnerable trees locally (Fig.~\ref{fig:3}b), each with small depth (Fig.~\ref{fig:3}d) \cite{Bakshy11,Goel2012Structure}. At the same time the stable adoption network (considering connections between all stable adopters at the time) has a giant connected component, indicating the emergence of a percolating stable cluster with size comparable to the largest adoption cluster (Fig.~\ref{fig:3}a, inset). These observations suggest a scenario for the evolution of the global adoption component different from earlier threshold models ~\cite{Watts2002Simple}. It appears that here multiple innovators adopt at different times and trigger local vulnerable trees (Fig.\ref{fig:1}b), which in turn induce a percolating component of connected stable nodes that holds the global adoption cluster together (Fig.\ref{fig:1}c). Consequently, in the structure of the adoption network primary triggering effects are important only locally, while external and secondary triggering mechanisms seem to be responsible for the emergence of global-scale adoption. 

To model the observed dynamics and explore the effect of immune nodes, we perform extensive numerical simulations of the threshold model with parameters determined directly from the data (see MM and SI). We use a network structure with empirical degree and threshold distributions and fix $p_n=0.00019$ as the constant rate of innovators, implying that the time scale of a Monte Carlo iteration in the model is 1 month. We measure the average size of the largest ($LC$) and second largest ($LC^{2nd}$) connected components of the background social network, and of the stable, vulnerable and global adoption networks, as a function of the fraction of immune nodes $r$. After $T=89$ iterations (matching the length of the real observation period) we identify three regimes of the dynamics (Fig.~\ref{fig:3}c): if $0<r<0.6$ (dark-shaded area) the spreading process is very rapid and evolves in a global cascade, which reaches most of the nodes of the shrinking susceptible network in a few iteration steps. About $10\%$ of adopters are connected in a percolating stable cluster, while vulnerable components remain very small in accordance with empirical observations. In the crossover regime $0.6<r<0.8$ (light-shaded area), the adoption process slows down considerably (Fig.~\ref{fig:2}d, lower panel), as stable adoptions become less likely due to the quenching effect of immune nodes. The adoption process becomes the slowest at $r_c=0.8$ (Fig.~\ref{fig:2}d, lower panel) when the percolating stable cluster falls apart, as demonstrated by a peak in the corresponding $LC^{2nd}$ curve (Fig.~\ref{fig:3}c, lower panel). Finally, around $r=0.9$ the adoption network becomes fragmented and no global diffusion takes place. We repeat the same calculations for another service and find qualitatively the same picture, but with the crossover regime shifted towards larger $r$ values due to the different parametrisation of the model process. Note that another possible reason for the slow adoption could be the time users wait between their threshold has been reached and actual adoption. We test for the effect of this potential scenario on the empirical curves but find no qualitative change in the dynamics (see SI).

We can use these calculations to estimate the only unknown parameter (the fraction $r$ of immune nodes in Skype) by matching the size of the largest component ($LC_{Net}$) between real and model adoption networks at time $T$. Empirically, this value is the relative size corresponding to the last point on the right-hand side of the distribution for $2011$ (Fig.~\ref{fig:3}a, main panel). The corresponding value in the model is $r = 0.73$ (dashed lines in Fig.~\ref{fig:3}c; also Fig.~\ref{fig:2}a and b), suggesting that the real adoption process lies in the crossover regime. The other analysed service turns out to lie right of the crossover regime, which explains its large innovator adoption rate and reduced size of stable and vulnerable adoption clusters (see SI).

To test the validity of the prediction of $r$ we perform three different calculations. First we measure the maximum relative growth rate of cumulative adoption and find a good match between model and data (Skype s3 and Model Skype s3 in Fig.~\ref{fig:1}a). In other words, the model correctly estimates the speed of the adoption process. Second, we measure the distribution $P(d)$ of depths of induced vulnerable trees (Fig.~\ref{fig:3}d). Finally, in order to verify earlier theoretical suggestions \cite{Singh2013Thresholdlimited}, we look at the correlation $\langle s_v \rangle (k)$ between the degree of innovators and the average size of vulnerable trees induced by them (Fig.~\ref{fig:3}e). We perform the last two measurements on the real data and in the model process for $r=0.6$ and $0.9$, as  well as for the predicted value $r=0.73$. In the case of $\langle s_v \rangle (k)$, we find a strong positive correlation in the data, explained partially by degree heterogeneities in the underlying social network, but surprisingly well emulated by the model. More importantly, although both quantities appear to scale with $r$, measures for the estimated $r$ value fit the empirical data remarkably well, confirming our earlier validation based on the matching of relative component sizes (for further discussion see SI).

\section*{Discussion}

Although some products and innovations diffusing in society may cover a large fraction of the population, their spreading tends to follow slow cascading patterns, the dynamics of which have been modelled before by simple diffusion models like that of Bass \cite{Bass1969}. However, this approach neglects threshold mechanisms that arguably drive the decision making of single individuals. On the other hand, threshold models study the conditions for cascades in global diffusion but do not address their temporal evolution, which is clearly a relevant factor in real-world adoption processes. These models are commonly used to predict rapid cascading patterns of adoption, which is a more realistic scenario for the spreading of information, opinions, or behavioural patterns but are not observed in the case of product or innovation diffusion where adoption requires additional efforts, e.g., free or paid registration. Here we provide a solution for this conundrum by analysing and modelling the worldwide spread of an online service in the techno-social communication network of Skype. Beyond the novel empirical evidence about heterogeneous adoption thresholds and non-linear dynamics of the adoption process, we identify two additional components necessary to introduce in the modelling of product adoption, namely: (a) a constant flow of innovators, which may induce rapid adoption cascades even if the system is initially out of the cascading regime; and (b) a fraction of immune nodes that forces the system into a quenched state where adoption slows down. These features are responsible for a critical structure of empirical adoption components that radically differs from previous theoretical expectations. We incorporate these mechanisms into a threshold model controlled by the rate of innovators and the fraction of immune nodes. The model is able to reproduce several pathways ranging from cascading behaviour to more realistic dynamics of innovation adoption. By constraining the model with empirically determined parameters, we provide an estimate for the real fraction of susceptible agents in the social network of Skype, and validate this prediction through correlated structural features matching empirical observations.

Our aim in this study was to provide empirical observations as well as methods and tools to model the dynamics of social contagion phenomena with the hope it will foster thoughts for future research. One possible direction would be the observation of the reported structure and evolution of the global adoption cluster in other systems similar to the ones studied in \cite{BorgeHolthoefer2011Structural,Dow2013Anatomy,Gruhl2004Information,Goel2012Structure,BorgeHolthoefer2013Cascading,Bakshy11}. Other promising directions could be the consideration of homophilic or assortative structural correlations, the evolving nature of the underpinning social structure as studied in \cite{Karsai2014Complex}, interpersonal influence, or the effects of leader-follower mechanisms on the social contagion process. Finally, we hope that the reported results may improve efficiency in the strategies of enhancing the diffusion of products and innovations, by shifting attention from the creation of short-lived perturbations to the sustenance of external input.

\subsubsection*{Competing interest statement}
The authors have no competing interests.

\subsubsection*{Authors' contributions statement}
M.K., G.I., R.K., K.K, and J.K designed the research and participated in writing the manuscript. R.K. and M.K. analysed the empirical data, G.I. made the analytical calculations, and M.K. and G.I. performed the numerical simulations.

\subsubsection*{Acknowledgements}
The authors gratefully acknowledge the support of M. Dumas, A. Saabas, and A. Dumitras from STACC and Microsoft/Skype Labs as well as constructive comments by J. Saram\"aki and T. N{\"a}si.

\subsubsection*{Funding statement}
G.I. acknowledges the Academy of Finland, and J.K. the CIMPLEX FET Open H2020 EU project for support. This research was partly funded by Microsoft/Skype Labs.

\section*{Material and Methods}

\section*{Data description}

We use a static representation of the Skype social network aggregated over 99 months between September 2003 and November 2011. We follow the adoption of the ``buy credit'' paid service for $89$ months starting from 2004, and the paid service ``subscription'' for $42$ months starting from 2008 (for further details about the network and service see SI). By considering the online social structure and adoption times, we identify users as innovator, vulnerable, or stable nodes based on the number $\Phi_k$ of adopting neighbours at the time of exposure. Thresholds are calculated as $\phi=\Phi_k/k$ for users with $k$ contacts. The adoption network is constructed by considering confirmed social links between users who adopted the service earlier than $t$. In order to avoid the effect of instantaneous group adoptions (evidently not driven by social influence), we only consider links between nodes who are neighbours in the underlying social network and whose adoption did not happen at the same time. Note that for the categorisation of nodes we use only the adoption time and the state of their peers, and thus real categories may differ slightly. For example, an innovator may appear as a vulnerable or stable node, even if its decision was not driven by social influence but some of its peers adopted earlier. To consider this bias we measure effective rates of adoption for the model process as well, just like for the empirical case (Fig.\ref{fig:1}) and section S3.

\section*{Maximum relative growth rate}

This measure is obtained by taking the maximum of the yearly adoption rate (yearly count of adoptions) normalised by the final observed adoption number of a given service. It characterises the maximum speed of adoption a service experienced during its history and takes values between 0 (no cascade) and 1 (instantaneous cascade). We repeat this measurement for the estimated number of registered users of Facebook, Twitter, and LinkedIn \cite{SocialMedia}, as well as for the number of active users of Skype and three paid Skype services. Adoption rates for Facebook, Twitter, and LinkedIn correspond to the period between 2006 and 2012, and for Skype and its services to the interval from release date until 2011.

\section*{Empirical parameter estimation}

We use the Skype data to directly determine all model parameters, apart from the fraction $r$ of immune nodes. To best approximate the degree distribution of the real network, after testing different candidate functions (see SI) we select a lognormal function $P(k) = e^{ -(\ln k-\mu_D)^2 / (2\sigma_D^2) } / (k\sigma_D\sqrt{2\pi})$ with parameters $\mu_D=1.2$ and $\sigma_D=1.39$ and minimum degree $\kmin = 1$, leading to the average degree $z = 8.56$. To account for finite-size effects in the model results for low $N$ (Fig.~\ref{fig:2}), we decrease $\mu_D$ slightly to obtain the same value of $z$ as in the real network.

The threshold distribution of each degree group collapses to a master curve after normalisation by using the scaling relation $P(\Phi_k,k)=k P (\Phi_k/k)$. This master curve can be well approximated by the lognormal distribution $P(\phi) = e^{ -(\ln \phi-\mu_T)^2 / (2\sigma_T^2) } / (\phi\sigma_T\sqrt{2\pi})$, with parameters $\mu_T=-2$ and $\sigma_T=1$ as determined by the empirical average threshold $w = 0.19$ and standard deviation $0.233$ (for further details see SI). We estimate a rate of innovators $p_n = 0.00019$ by fitting a constant function to $R_i(t)$ for $t > 2\tau$ (Fig.~\ref{fig:1}f). The fit to $\pn$ also matches the time scale of a Monte Carlo iteration in the model to 1 month. Model results (Fig.~\ref{fig:3}d and e) are calculated with $r = 0.73$ and $p_n = 0.00019$. Simulation results in Fig.~\ref{fig:3}c (d and e) are averaged over $100$ configuration-model networks of size $N=10^5$ ($10^6$) after $T=89$ iterations, matching the length of the observation period in Skype.

\section*{Model description}

We characterise the static social network by the extended distribution $\Pk$, where $\Pk = r P(k)$ for $c = 0$ and $\Pk = (1 - r) P(k) P(c)$ for $c > 0$. Non-immune, susceptible nodes with property vector $\kvec$ adopt spontaneously at a constant rate $\pn$, else they adopt only if a fraction $\phi_c$ of their $k$ neighbours have adopted before. These rules are condensed in the probability $\Fkm dt$ that a node will adopt in a small time interval $dt$, given that $m$ of its neighbours are already adopters,
\begin{equation}
\label{eq:thresRule}
\Fkm =
\begin{cases}
\pr & \text{if} \quad m < k \phi_c \\
1 & \text{if} \quad m \geq k \phi_c
\end{cases}, \quad \forall m \; \text{and} \; k, c \neq 0,
\end{equation}
with $F_{(k,0),m} = 0$ $\forall k, m$ and $F_{(0,c),0} = \pr$ $\forall c \neq 0$ (for immune and isolated nodes, respectively). The dynamics of adoption is well described by an AME for the fraction $\skm(t)$ of $\kvec$-nodes that are susceptible at time $t$ and have $m=0,\ldots,k$ adopting neighbours~\cite{porter2014,gleeson2013binary,gleeson2011high},
\begin{equation}
\label{eq:AMEsThres}
\dskm = -\Fkm \skm -\bs (k - m) \skm + \bs (k - m + 1) \skmo,
\end{equation}
where $\bs(t) = \frac{\sumk \Pk \summ (k - m) \Fkm \skm(t)}{\sumk \Pk \summ (k - m) \skm(t)}$. To reduce the dimensionality of Eq.~(\ref{eq:AMEsThres}) we consider the ansatz $\skm = \Bkm (\nu) e^{-\pr t}$ for $m < k\phi_c$, leading to the condition $\dot{\nu} = \bs (1 - \nu)$. With $\rho = 1 - \sumk \Pk \summ \skm$ and some algebra, this condition is reduced to Eq.~(\ref{eq:reducedAMEs}) (see SI).

\newpage

\section*{\LARGE{Supplementary Information}}


\section{Detailed data description}
\label{sec:ddescr}

This study has been conducted on a dataset of the social network of Skype. The centrepiece of the dataset is the \emph{contact network}, where nodes represent users and edges between pairs of users exist if they are in each other's contact lists. A user's contact list is composed of \emph{friends}. If user $u$ wants to add another user $v$ to his/her contact list, $u$ sends $v$ a contact request, and the edge is established at the moment $v$ approves the request (or not, if the contact request is rejected). Each edge is labelled with a time stamp indicating the moment the contact request was approved. As the underpinning social structure we consider the static representation of the Skype social network, aggregated for $99$ months between September 2003 and November 2011. The largest connected component of this structure includes roughly 510 million users and 4.4 billion edges.

As the chosen service evolving on the Skype network, we follow how users purchase ``credits'' for calling phones. For each user, the dataset includes the date when he/she first adopted the paid product ``buy credit'' (first credit purchase, for all purposes). We select this service since its lifetime of $89$ months is considerably long (it was introduced in 2004), and it can be adopted by registered Skype users only. This way the aggregated Skype network provides a complete description of the mediating social structure, which allows us to calculate the correct degree and adoption threshold for all individuals. To make additional observations and to further test our model, we perform calculations on a second paid service called ``subscription'', which was introduced in April 2008, lasts for over $42$ months, and can also be adopted by registered Skype users only. Results regarding this service are presented in Section \ref{sec:addserv}.

By considering the online social structure and the adoption times we identify users as innovator, vulnerable, or stable nodes based on the number $\Phi_k$ of adopting neighbours at the time of exposure. Thresholds are calculated as $\phi=\Phi_k/k$ for users with $k$ contacts. The adoption network is constructed by considering confirmed social links between users who adopted the service earlier than the time of observation $t$. In order to avoid the effect of instantaneous group adoptions (evidently not driven by social influence), we only consider links between nodes who are neighbours in the underlying social network and whose adoption did not happen at the same time. Note that for the categorization of nodes we use only the adoption time and the state of their peers, and thus `real' categories may differ slightly. For example, an innovator may appear as a vulnerable or stable node, even if its decision was not driven by social influence but some of its peers adopted earlier. To consider this bias we also measure `effective' rates of adoption for the model process, just like for the empirical case (Fig.1, main text) and section S3.

The dataset does not include identity information. All usernames are anonymized and there is no way of inferring a user's identity solely from the profile. The dataset does not contain any information about interpersonal interactions, apart from the contact list.

\section{Empirical determination of model parameters}
\label{sec:pars}

Parameters in the model are the rate of innovators $p_{n}$, the degree distribution $P(k)$, the threshold distribution $P(\phi)$, and the fraction of immune nodes $r$. Other than $r$, all of them can be estimated from the data as follows.

\subsection{Rate of innovators}

As discussed in the main text, the rate of spontaneous adoption saturates approximately to a constant value after an initial transition period, which allows us to determine the rate of innovators by fitting a constant function on the curve after time $2\tau$. We estimate this rate to be $p_{n} = 0.00019$, as demonstrated in Fig.~S\ref{fig:NullModelRate}a where the dashed line assigns the fitted constant function.

\subsection{Degree distribution}
\label{sec:degDistrFit}

\begin{figure}[t] \centering
\subfigure[]{
  \includegraphics[width=76mm]{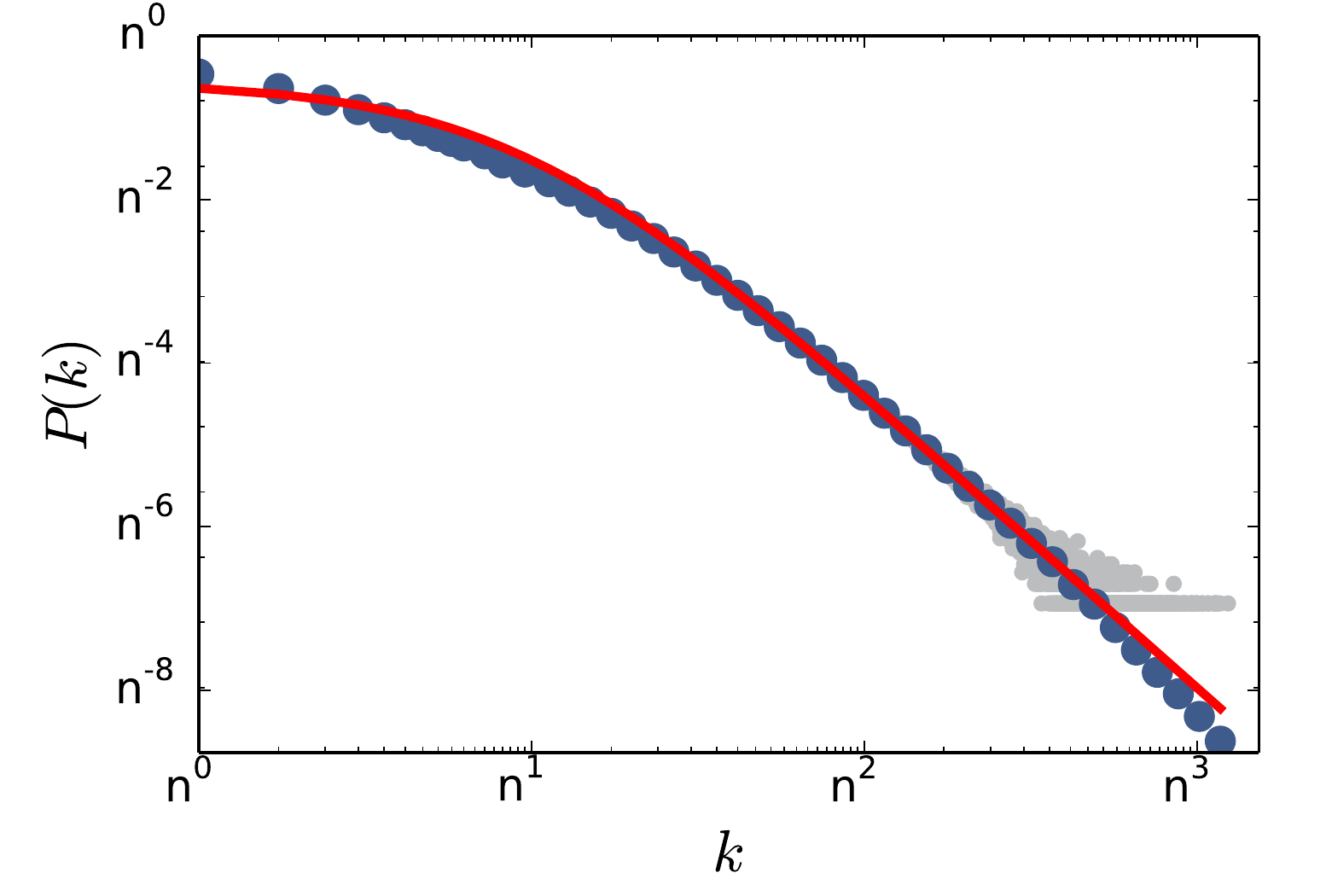}
}
\subfigure[]{
  \includegraphics[width=76mm]{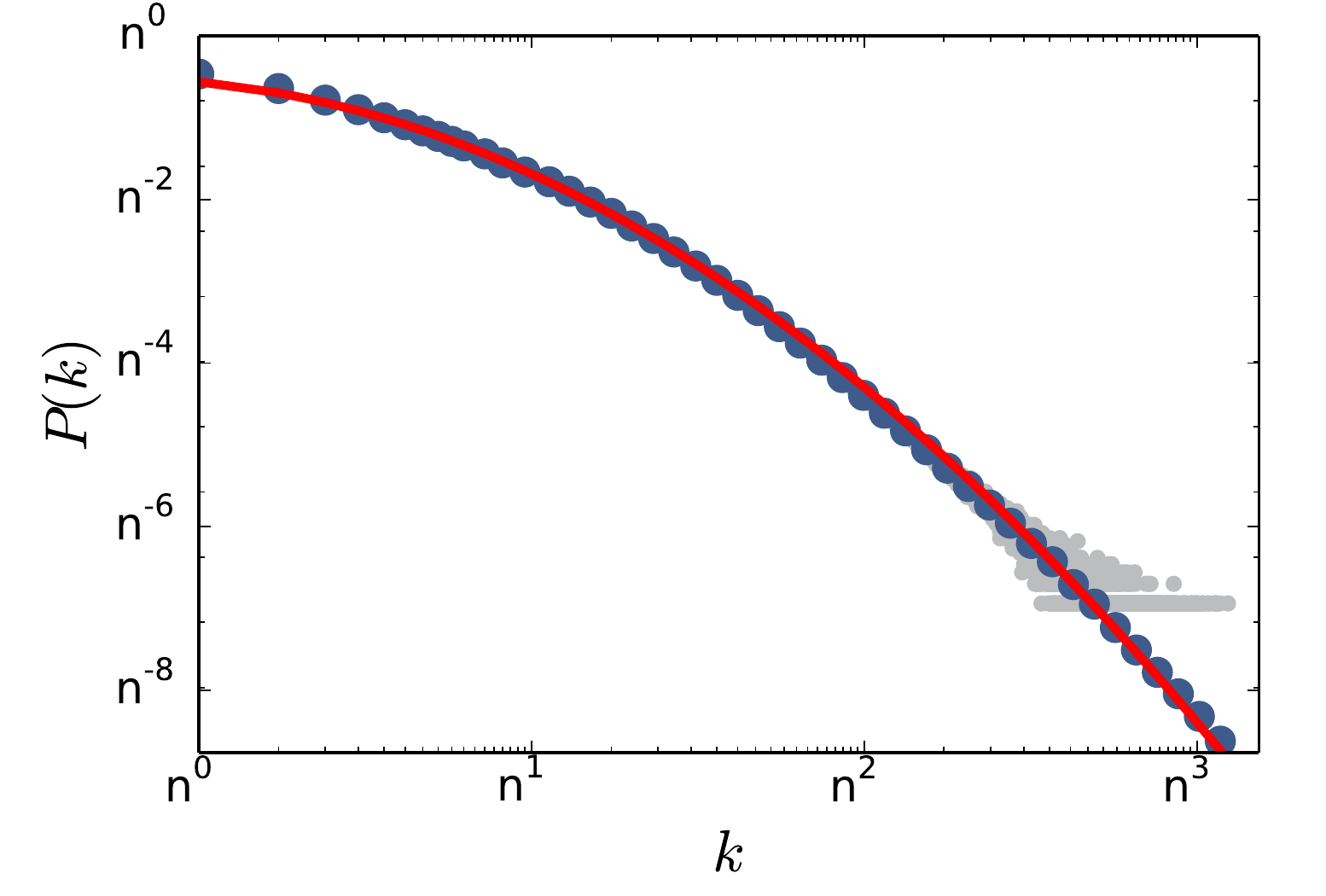}
} \\
\caption{\small{\bf Fitted degree distribution.} {\bf (a)} Empirical degree distribution $P(k)$ fitted with a shifted power-law distribution function [Eq.~(\ref{eq:shSF})] with parameters described in the text. {\bf (b)} $P(k)$ fitted by a lognormal distribution function [Eq.~(\ref{eq:logn})] with parameters determined in the text. Grey symbols are the degree distribution, blue symbols the log-binned distribution, and solid lines the fitted analytical distribution.}
\label{fig:degdistr}
\end{figure}

\begin{figure}[t] \centering
\subfigure[Empirical adoption rates]{
	\includegraphics[width=76mm]{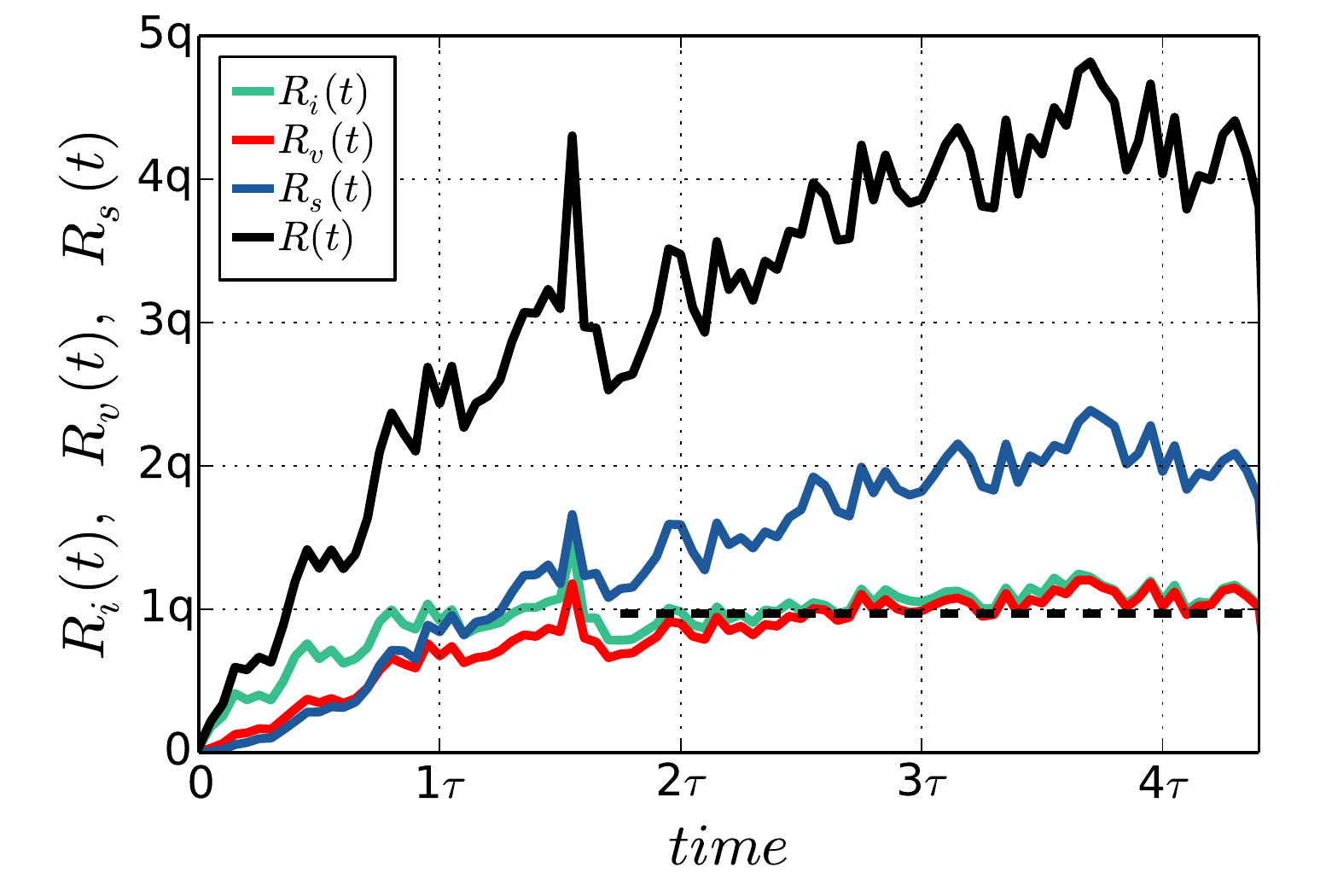}
}
\subfigure[Null model adoption rates]{
	\includegraphics[width=76mm]{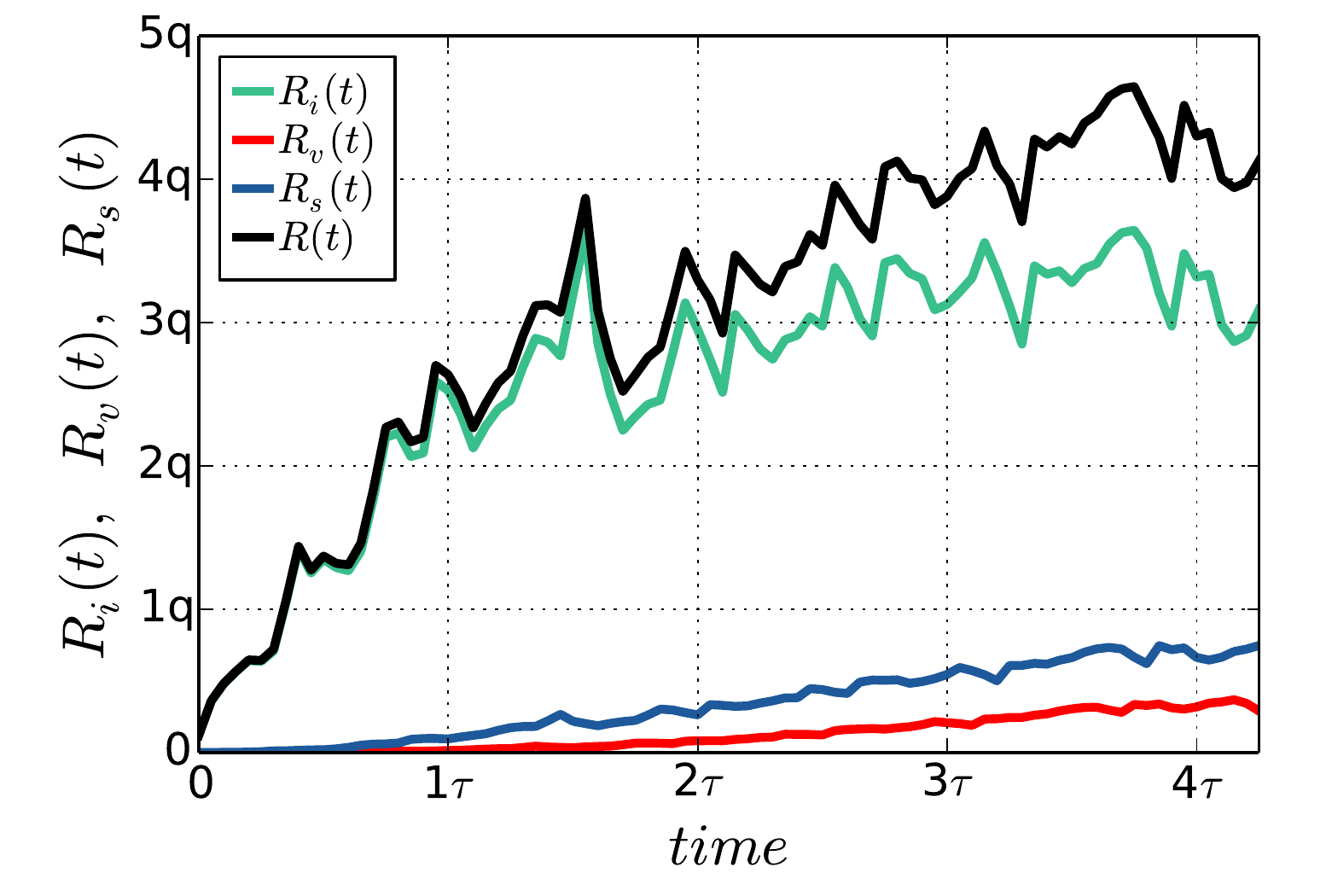}
}
\caption{\small{\bf Adoption rates in the original and null model processes.} Adoption rates for innovator (green), vulnerable (red), and stable (blue) nodes as a function of time. {\bf (a)} Empirical rates where adoptions appear in the original order. The dashed line assigns a fitted constant function to estimate the innovator adoption rate. {\bf (b)} Null model rates where times of adoption are randomly shuffled. Here $q$ and $\tau$ are arbitrary constant values.}
\label{fig:NullModelRate}
\end{figure}

Degrees in the aggregated Skype network are broadly distributed with a fat tail corresponding to strong degree heterogeneities. To characterize this distribution analytically we select two candidate distribution functions. The first is a shifted power-law distribution function of the form,
\begin{equation}
P(k)=\frac{\gamma-1}{C+k_{min}}\left( \frac{C+k}{C+k_{min}} \right)^{-\gamma} \hspace{.2in} \mbox{for} \hspace{.2in}	k_{min}\leq k,
\label{eq:shSF}
\end{equation}
where $k$ denotes the degree, $\gamma$ is the power-law exponent scaling the tail of the distribution, and $k_{min}$ is the minimum degree (in our case 1). $C$ is a constant scaling the shift of the distribution, which can be determined as $C=z(\gamma-2)-k_{min}(\gamma-1)$ since we know the average degree $z = 8.56$ of the empirical network. This way our only free parameter during the fit is the degree exponent $\gamma$. After fitting this function by using the non-linear least-square method, we obtain a relatively good match with the empirical distribution (Fig.~S\ref{fig:degdistr}a) for exponent $\gamma=3.61$.

Our second candidate function is a lognormal distribution function of the form,
\begin{equation}
P(k)=\frac{1}{k\sigma_D\sqrt{2\pi}}e^{-\frac{(\ln k-\mu_D)^2}{2\sigma_D^2}} \hspace{.2in} \mbox{for} \hspace{.2in}	k_{min}\leq k,
\label{eq:logn}
\end{equation}
where $\mu_D$ and $\sigma_D$ are the scaling parameters. After fitting this function by using the non-linear least square method with two free parameters ($\mu_D$ and $\sigma_D$), we obtain an excellent fit with the empirical distribution for parameters $\mu_D=1.2$ and $\sigma_D=1.39$.

To select the best candidate function, we calculate the corresponding Jensen-Shannon ($JS$) divergence values \cite{Lin1991} between the empirical and fitted distributions. As a result we find that for the shifted power-law function the best fit provides $JS=0.0257$, while for the lognormal distribution  we get $JS=0.0051$. Thus we select the lognormal distribution as the best analytical function describing the degree distribution of the empirical network.

\subsection{Threshold distribution}
\label{sec:thrDistr}

The adoption threshold $\phi$ of a node is defined as $\phi = \Phi_k / k$, i.e. the fraction of adopting neighbours that trigger the adoption of the central node. Therefore it can only take certain fractional values determined by the degree $k$. Although thresholds are defined as a fraction, by considering nodes of the same degree we can focus on the integer threshold $\Phi_k$, defined as the number of a node's neighbours who have adopted the service earlier.

In our method we first group nodes of the same degree, record their integer thresholds, and then calculate the threshold distribution for each degree group, as shown in the main text (Fig.~1e, inset). These distributions collapse to a master curve after normalization by using the scaling relation $P(\Phi_k,k)=k P (\Phi_k/k)$ (Fig.~1e, main panel). Moreover, this master curve can be well approximated by a lognormal distribution of the form,
\begin{equation}
P(\phi)=\frac{1}{\phi \sigma_T\sqrt{2\pi}}e^{-\frac{(\ln\phi-\mu_T)^2}{2\sigma_T^2}},
\label{eq:thrLogn}
\end{equation}
where $\mu_T=-2$ and $\sigma_T=1$, as determined by the empirical average threshold $w = 0.19$ and standard deviation (STD) $0.233$.

These findings indicate that although individual thresholds are strongly determined by degree, their distribution is degree-invariant, suggesting that the fraction of adopting friends rather than its absolute number is relevant during the service adoption process. The estimated empirical values of parameters are summarized in Table \ref{table:pars}. 

\begin{table}[h]
\begin{center}
\begin{tabular}{|c||c|c|c||c|c|c|c|}
\hline 
$p_{n}$ & $\langle k \rangle$ & $\mu_D$ & $\sigma_D$ & $w$ & $STD(\phi)$ & $\mu_T$ & $\sigma_T$  \\ \hline \hline 
$0.00019$ & $8.56$ & $1.2$  & $1.39$ & $0.19$ & $0.233$ & $-2$ & $1$ \\
\hline 
\end{tabular} 
\caption{\small Estimated empirical parameters for service ``buy credit''.}
\label{table:pars}
\end{center}
\end{table}

\section{Social influence - null model study}
\label{sec:sinf}

Studies of social contagion phenomena assume that social influence is responsible for the correlated adoption of connected people. However, an alternative explanation for the observed correlated adoption patterns is homophily: a link creation mechanism by which similar egos get connected in a social structure. In the latter case, the correlated adoption of a connected group of people would be explained by their similarity and not necessarily due to social influence. Homophily and influence are two processes that may simultaneously play a role during the adoption process. Nevertheless, distinguishing between them on the individual level is very difficult using our or any similar datasets \cite{Shalizi2011}. Fortunately, at the system level one may decide which process is dominant in the empirical data. To do that first we need to elaborate on the differences between these two processes.

Influence-driven adoption of an ego can happen once one or more of its neighbours have adopted, since then their actions may influence the decision of the central ego. Consequently, the time ordering of adoptions of the ego and its neighbours matters in this case. Homophily-driven adoption is, however, different. Homophily drives social tie formation such that similar people tend to be connected in the social structure. In this case connected people may adopt because they have similar interests, but the time ordering of their adoptions would not matter. Therefore, it is valid to assume that adoption could evolve in clusters due to homophily, but these adoptions would appear in a random order.

To test our hypothesis we define a null model where we take the adoption times of each adopter and shuffle them randomly among all adopting egos. This way a randomly selected time is assigned to each adopter, while the adoption rate and the final set of adopters remain the same. Moreover, this procedure only destroys correlations between adoption events induced by social influence, but keeps the social network structure and node degrees unchanged. In this way, during the null model process the same egos appear as adopters, but the time series of adoption may in principle change (or not), corresponding to social influence (or homophily) as a dominant factor during the adoption process. If adoption is mostly driven by homophily, the rates of adoption would not change considerably beyond statistical fluctuations. On the other hand, if social influence plays a role in the process, rates of adoption in the null model should be very different from the empirical curves, implying that the time ordering of events matters in the adoption process. In this case, the rate of innovators should be higher than in the empirical data, since nodes that are in the adoption cluster originally but not directly connected would have a greater chance to appear as innovators, due to a random adoption time that is not conditional to the time ordering of the adopting neighbours.

After calculating the adoption rates of different user groups in the shuffled null model sequence we observe the latter situation: the rate of innovators becomes dominant, while the rate of stable and vulnerable adoptions drops considerably as they appear only by chance. This suggests that the temporal ordering of adoption events matters a lot in the evolution of the observed adoption patterns, and thus social influence may play a strong role here. Of course one cannot decide whether influence is solely driving the process or homophily has some impact on it; in reality it probably does to some extent. However, we can use this null model measure to demonstrate the presence and importance of the mechanism of influence during the adoption process.

\section{Threshold model}
\label{sec:thresModel}

\subsection{Model description}
\label{ssec:modDesc}

This model emulates the rise and temporal evolution of system-wide adoption cascades in complex social networks~\cite{watts2002,singh2013,Ruan2015}. Note that this model has been introduced in \cite{Ruan2015}, where its general scaling behaviour has been explored. In a system of fixed size, a node has social interactions with $k$ other agents and is characterized by a continuous adoption threshold $\phi$. When faced with the prospect of adopting a given innovation, product, or fad, susceptible individuals adopt spontaneously with rate $\pn$. Otherwise, the node adopts if at least a fraction $\phi$ of its $k$ neighbours have adopted before (the so-called `threshold rule'). Moreover, a fraction $r$ of the system is `immune' to the innovation, in the sense that these agents never adopt regardless of their values of $k$ and $\phi$. The distributions of degrees and thresholds, $P(k)$ and $P(\phi)$ (as well as the values of $\pn$ and $r$), thus determine the average topological state and dynamical evolution of the system.

The model may be implemented numerically via a Monte Carlo simulation of the rules described above in a system of size $N$. Here, the dynamical state of the system is determined by the adoption state (0 or 1) of all nodes, which change in asynchronous random order in a series of time steps. Once an agent adopts and its state changes from 0 to 1, it remains so for the rest of the dynamics, thus ensuring a frozen final state for the finite system where no more adoptions arise. Each time step consists of $N$ node updates: In each node update, a randomly selected node (non-immune and in state 0) adopts spontaneously with probability $\pr = \pn / (1 - r)$ \footnote{We define $\pr = 1$ for $\pn > 1 - r$.}; else it adopts only if the threshold rule is satisfied. The rescaled rate $\pr$ is necessary if we wish to obtain a rate $\pn$ of innovators in early times of the dynamics, regardless of the value of $r$. Finally, we assume that agents with $k=0$ receive no social influence (for any value of $\phi$), and therefore can only adopt spontaneously. We will now explore this dynamics with numerical simulations and a rate equation formalism.

\subsection{Stochastic binary-state dynamics}
\label{ssec:stocDynam}

Here we extend an approximate master equation (AME) formalism for stochastic binary-state dynamics as developed recently by Gleeson~\cite{porter2014,gleeson2013,gleeson2008,gleeson2011}. In a stochastic binary-state dynamics, each node in the network can take one of two possible states (susceptible or adopter in the language of innovation adoption) at any point in time, and state-switching happens randomly with probabilities that only depend on the current state of the updating agent and on the states of its neighbours. This general definition includes the threshold model described above as a special case. Such formalism considers configuration-model networks, that is, an ensemble of networks specified by the degree distribution $P(k)$ but otherwise maximally random~\cite{newman2010}.

All relevant properties used to describe a node are included in the vector $\kvec = (k, c)$, where $k = k_0, k_1, \ldots k_{M-1}$ is the degree of the node and $c = 0, 1, \ldots, M$ a dummy variable that labels its `type', i.e. any other property that characterizes the node apart from its degree. In the case of our threshold model, $c = 0$ is the type of the fraction $r$ of immune nodes, while $c \neq 0$ labels the type of all non-immune nodes with given threshold $\phi_c$. The various values of $c \neq 0$ correspond then to different adoption thresholds $\phi_c$. The integer $M$ is the maximum number of degrees/types considered in the AME framework, which can be increased to improve the accuracy of the analytical approximation at the expense of speed in its numerical computation\footnote{Explicitly, rather than using $k_0 = k_{\text{min}}$, $k_{M-1} = N - 1$ and $M = N - k_{\text{min}}$ (i.e, considering all possible degrees in the empirical/simulated network), we take a small $k_0 > k_{\text{min}}$ and large $k_{M-1} < N - 1$, $M < N - k_{\text{min}}$, with the other $M - 2$ degree values equidistantly distributed between $k_0$ and $k_{M-1}$, thus disregarding some degrees and gaining speed in the computation of the AMEs. Similarly, the $M$ threshold values corresponding to nonzero types are uniformly distributed in the open interval $(0, 1)$.}. Any pair of nodes with identical values of $\kvec$ are considered equivalent in this level of description, forming a node class with the same average dynamics. Moreover, $P(k)$ and $P(\phi)$ can be generalized to the joint distribution $\Pk$ giving the probability that a randomly selected node has property vector $\kvec$ (i.e. degree $k$ and type $c$). Here it is useful to define $P(c)$ as the distribution of all non-zero types, $c = 1, \ldots, M$. If degrees and thresholds are chosen independently, like in our model, then $\Pk = r P(k)$ for $c = 0$ and $\Pk = (1 - r) P(k) P(c)$ for $c > 0$. The distribution $P(c)$ is, in other words, a discrete, rescaled version of the continuous threshold distribution $P(\phi)$.

In the language of innovation adoption, the dynamics of a node is determined by the number $m = 0, 1, \ldots k$ of its neighbours that have already adopted when the node is deciding whether to adopt or not. During a small time interval $dt$, a susceptible node (in state 0) adopts with probability $\Fkm dt$, while an adopter (in state 1) becomes susceptible with probability $R_{\kvec, m} dt$. The functions $\Fkm$ and $R_{\kvec, m}$, known as infection and recovery rates, respectively, determine the temporal evolution of the node class $\kvec$. In the particular case of threshold models, a so-called monotone dynamics, $R_{\kvec, m} = 0$ $\forall\, \kvec, m$ (since no adopters become susceptible again). As for $\Fkm$, the rules of spontaneous and threshold adoption imply,
\begin{equation}
\label{eq:thresRule}
\Fkm =
\begin{cases}
\pr & \text{if} \quad m < k \phi_c \\
1 & \text{if} \quad m \geq k \phi_c
\end{cases}, \quad \forall m \; \text{and} \; k, c \neq 0,
\end{equation}
that is, a node adopts the innovation either spontaneously with rate $\pr$, or with probability 1 if its number of adopting neighbours equals or exceeds the integer threshold $\Phi_k = \lceil k \phi_c \rceil$. Immune nodes ($c = 0$) have an infection rate of $F_{(k,0),m} = 0$ $\forall k, m$, while for isolated nodes ($k = 0$) $F_{(0,c),0} = \pr$ $\forall c \neq 0$. In other words, immune nodes never adopt, and isolated nodes can only adopt spontaneously. We note that $\Fkm$ is written in terms of $\pr = \pn / (1 - r)$, not $\pn$, in order to counter the trivial decrease in the rate of spontaneous adoption for non-zero $r$.

Let us now turn to the rate equations for our threshold model, called AMEs in the formalism by Gleeson. We denote by $\skm(t)$ the fraction of $\kvec$-class nodes that are susceptible at time $t$ and have $m$ adopting neighbours. Therefore, the fraction of agents with property vector $\kvec$ that are adopters at time $t$ is $\pk(t) = 1 - \sum_{m=0}^k \skm (t)$, and the fraction of adopters in the system is $\rho (t) = \sumk \Pk \pk(t)$. Here, the sum over classes means a sum over all degrees and types, i.e. $\sumk \bullet = \sum_k \sum_c \bullet$. Assuming a monotone dynamics ($R_{\kvec, m} = 0$), the AMEs for $\skm$ can be written as~\cite{porter2014,gleeson2013,gleeson2011},
\begin{equation}
\label{eq:AMEsThres}
\frac{d \skm}{dt} = -\Fkm \skm -\bs (k - m) \skm + \bs (k - m + 1) \skmo,
\end{equation}
where $m = 0, \ldots, k$, $s_{\kvec, -1} \equiv 0$, $\Fkm$ follows Eq.~(\ref{eq:thresRule}), and $\bs(t)$ (the rate at which edges between pairs of susceptible nodes transform to edges between a susceptible agent and an adopter) is given by,
\begin{equation}
\label{eq:rateBs}
\bs(t) = \frac{\sumk \Pk \summ (k - m) \Fkm \skm(t)}{\sumk \Pk \summ (k - m) \skm(t)}.
\end{equation} 
If at time $t = 0$ we randomly choose a fraction $\rho (0) = \sumk \Pk \pk(0)$ of nodes as seed for the adoption process, the initial conditions for Eq.~(\ref{eq:AMEsThres}) are $\skm (0) = [1 - \pk(0)] \Bkm [\rho(0)]$, with $\pk(0)$ the initial fraction of adopters in class $\kvec$ and $\Bkm$ a binomial factor,
\begin{equation}
\label{eq:BinomFac}
\Bkm(\rho) = \binom{k}{m} \rho^m (1 - \rho)^{k - m}.
\end{equation}

The solution $\skm(t)$ of the AME system in Eq.~(\ref{eq:AMEsThres}) provides a very accurate description of the dynamics of our model, yet its dimension  (i.e. number of equations to solve) grows quadratically with the number of degrees and linearly with the number of threshold values considered. Fortunately, the AMEs for our model can be mapped to a reduced-dimension system with a derivation similar to the one used by Gleeson in the case of the Watts threshold model~\cite{watts2002,singh2013}.

\subsection{Reduced-dimension AMEs}
\label{ssec:redAMEs}

To reduce the dimension of Eq.~(\ref{eq:AMEsThres}), we need to consider system-wide quantities that are more aggregated than $\skm$. One of them is the probability that a randomly chosen node is an adopter, $\rho(t) = 1 - \sumk \Pk \summ \skm (t)$, i.e. the fraction of adopters in the network. The other one is the probability that a randomly chosen neighbour of a susceptible node is an adopter, $\nu(t) = \sumk \Pk \summ m \skm(t) / \summ k \skm(t)$.

We start by proposing an exact solution for the AME system in terms of the following ansatz,
\begin{equation}
\label{eq:AMEansatz}
\skm(t) = [1 - \pk(0)] \Bkm [\nu(t)] e^{-\pr t}
\quad \text{for} \; m < k\phi_c  \; \text{and} \; c \neq 0,
\end{equation}
and $s_{(k,0),m} = \Bkm(\nu)$ for $c = 0$, where $\Bkm$ follows Eq.~(\ref{eq:BinomFac}). The meaning of the ansatz in Eq.~(\ref{eq:AMEansatz}) is quite intuitive and considers two processes. First, a susceptible agent with degree $k$ and $m$ adopting neighbours is not selected as part of the initial adoption seed with probability $1 - \pk(0)$ and is connected to $m$ adopters with the binomially distributed probability $\Bkm(\nu)$. Second, for $m < k\phi_c$ a susceptible node does not fulfill the threshold rule and can only adopt spontaneously with probability $e^{-\pr t}$, since the system is progressively been filled by adopters. Considering these two processes as independent we end up with the product in Eq.~(\ref{eq:AMEansatz}). Finally, since immune nodes do not adopt and are distributed randomly over the network, $s_{(k,0),m}$ is determined only by a binomial factor.

The next step is to insert the ansatz~(\ref{eq:AMEansatz}) into the AME system~(\ref{eq:AMEsThres}) and derive a set of differential equations for the aggregated quantities $\rho$ and $\nu$. Taking the time derivative $\dskm$ of Eq.~(\ref{eq:AMEansatz}) (i.e. the left-hand side of Eq.~(\ref{eq:AMEsThres})) we get,
\begin{equation}
\label{eq:ansatzINames1}
\dskm = \left( \left[ \frac{m}{\nu} - \frac{k-m}{1-\nu} \right] \dot{\nu} - \pr \right) \skm.
\end{equation}
Then, we use the threshold rule~(\ref{eq:thresRule}) for $m < k\phi_c$, the ansatz~(\ref{eq:AMEansatz}) and the binomial identity,
\begin{equation}
\label{eq:binomIdent}
\Bkmo(\nu) = \frac{1-\nu}{\nu} \frac{m}{k-m+1} \Bkm(\nu),
\end{equation}
in the right-hand side of Eq.~(\ref{eq:AMEsThres}) to obtain,
\begin{equation}
\label{eq:ansatzINames2}
-\Fkm \skm -\bs (k - m) \skm + \bs (k - m + 1) \skmo = \left[ -\pr + \bs \left( m - k + \frac{1-\nu}{\nu}m \right) \right] \skm.
\end{equation}
Equating Eqs.~(\ref{eq:ansatzINames1}) and~(\ref{eq:ansatzINames2}) as in the AME system~(\ref{eq:AMEsThres}) leads to,
\begin{equation}
\label{eq:condNu}
\dot{\nu} = \bs (1 - \nu),
\end{equation}
a condition on $\nu$ so that the ansatz~(\ref{eq:AMEansatz}) is a solution of Eq.~(\ref{eq:AMEsThres}). This differential equation has the initial condition $\nu(0) = \rho(0)$, obtained by evaluating Eq.~(\ref{eq:AMEansatz}) at $t = 0$ and comparing with the expression $[1 - \pk(0)] \Bkm [\rho(0)]$, which corresponds to a random distribution of initial adopters among $\kvec$ classes. Furthermore, by assuming a (yet to be determined) function $g(\nu, t)$ such that $\dot{\nu} = g(\nu, t) - \nu$, Eq.~(\ref{eq:condNu}) reduces to,
\begin{equation}
\label{eq:condBeta}
\bs = \frac{g(\nu, t) - \nu}{1 - \nu}.
\end{equation}

Now, we consider the following general result derived by Gleeson in~\cite{gleeson2013} (Eqs.~(F6)--(F10) therein),
\begin{equation}
\label{eq:GleesonEq}
\sumk \Pk \summ (k - m) \skm = z (1 - \nu)^2,
\end{equation}
with $z = \sum_k k P(k)$ the average degree in the network. Eq.~(\ref{eq:GleesonEq}) is valid for functions $\skm$ and $\nu$ that satisfy Eqs.~(\ref{eq:AMEsThres}) and (\ref{eq:condNu}) respectively, for any $\Fkm$ and random initial conditions on $\skm$ and $\nu$, and is thus applicable in our case. Our goal here is to use Eq.~(\ref{eq:GleesonEq}) to find an expression for $g(\nu)$ and therefore write the differential equation~(\ref{eq:condNu}) explicitly. Noting that the left-hand side of Eq.~(\ref{eq:GleesonEq}) is the denominator in the definition~(\ref{eq:rateBs}) of $\bs$ and that $F_{(k,0),m} = 0$ (i.e. immune nodes do not adopt), Eq.~(\ref{eq:rateBs}) gives,
\begin{align}
\label{eq:betaExpl1}
\bs &= \frac{1 - r}{z (1 - \nu)^2} \left[ \pr \sumkc P(k) P(c) \summLess (k - m) \skm + \sumkc P(k) P(c) \summMore (k - m) \skm \right] \nonumber\\
&= \frac{1}{z (1 - \nu)^2} \left[ \sumk \Pk \summ (k - m) \skm - r \sum_k P(k) \summ (k - m) s_{(k,0),m} \right. \nonumber\\
&\quad \left. - (1 - r) (1 - \pr) \sumkc P(k) P(c) \summLess (k - m) \skm \right],
\end{align}
where we have written $\Pk$ explicitly as $\Pk = r P(k)$ for $c = 0$ and $\Pk = (1 - r) P(k) P(c)$ for $c > 0$, in order to notice the dependence on $r$. Then, we insert the ansatz~(\ref{eq:AMEansatz}) (with its special case $s_{(k,0),m} = \Bkm(\nu)$ for immune nodes), as well as the identities $(k - m) \Bkm(\nu) = k (1 - \nu) \Bkom(\nu)$ and $\summLess \Bkom(\nu) = 1 - \summMore \Bkom(\nu)$ to obtain,
\begin{align}
\label{eq:betaExpl2}
\bs &= \frac{1}{1 - \nu} \Bigg( (1 - r) \Bigg[ 1 - (1 - \pr) e^{-\pr t} \Bigg. \Bigg. \nonumber\\
&\quad \left. \left. + (1 - \pr) e^{-\pr t} \sumkc \frac{k}{z} P(k) P(c) \left( \pk(0) + [1 - \pk(0)] \summMore \Bkom(\nu) \right) \right] - \nu \right).
\end{align}

A comparison of Eqs.~(\ref{eq:condBeta}) and~(\ref{eq:betaExpl2}) gives us the following expression for $g(\nu, t)$,
\begin{equation}
\label{eq:gFactor}
g(\nu, t) = (1 - r) \left( \ft + (1 - \ft) \sumkc \frac{k}{z} P(k) P(c) \left[ \pk(0) + [1 - \pk(0)] \sum_{m \geq k\phi_c} \Bkom(\nu) \right] \right),
\end{equation}
where we define $\ft$ as $\ft = 1 - (1 - \pr) e^{-\pr t}$. Thus, the AME system~(\ref{eq:AMEsThres}) gets reduced to the differential equation $\dot{\nu} = g(\nu, t) - \nu$, with $g(\nu, t)$ given explicitly by Eq.~(\ref{eq:gFactor}).

Even though the equation $\dot{\nu} = g(\nu, t) - \nu$ is closed and in this sense equivalent to Eq.~(\ref{eq:AMEsThres}), we can also derive the corresponding equation for $\rho$, since we are mainly interested in the temporal evolution of the fraction of adopters in the network. From the definition of $\rho$ and Eq.~(\ref{eq:AMEsThres}) we have,
\begin{align}
\label{eq:rhoDeriv1}
\dot{\rho} = - \sumk \Pk \summ \dskm &= \sumk \Pk \summ \Fkm \skm \nonumber\\
&\quad + \bs \sumk \Pk \summ \big[ (k - m) \skm - (k - m + 1) \skmo \big],
\end{align}
where the second term in the right-hand side telescopes to zero. Then, we use an algebraic manipulation similar to that of Eqs.~(\ref{eq:betaExpl1}) and~(\ref{eq:betaExpl2}) to obtain,
\begin{align}
\label{eq:rhoDeriv2}
& \sumk \Pk \summ \Fkm \skm = (1 - r) \left( \pr \sumkc P(k) P(c) \summLess \skm + \sumkc P(k) P(c) \summMore \skm \right) \nonumber\\
&= (1 - r) \left( 1 - (1 - r) (1 - \pr) \sumkc P(k) P(c) \summLess \skm \right) - \rho \nonumber\\
&= (1 - r) \left( \ft + (1 - \ft) \sumkc P(k) P(c) \left[ \pk(0) + [1 - \pk(0)] \sum_{m \geq k\phi_c} \Bkm(\nu) \right] \right) - \rho.
\end{align}
In this way, Eqs.~(\ref{eq:rhoDeriv1}) and~(\ref{eq:rhoDeriv2}) can be rewritten as $\dot{\rho} = h(\nu, t) - \rho$, where,
\begin{equation}
\label{eq:hFactor}
h(\nu, t) = (1 - r) \left( \ft + (1 - \ft) \sumkc P(k) P(c) \left[ \pk(0) + [1 - \pk(0)] \sum_{m \geq k\phi_c} \Bkm(\nu) \right] \right).
\end{equation}

Joining all of these results, the AME system~(\ref{eq:AMEsThres}) gets reduced to the system of two ordinary differential equations,
\begin{subequations}
\label{eq:reducedAMEs}
\begin{align}
\dot{\rho} &= h(\nu, t) - \rho, \\
\dot{\nu} &= g(\nu, t) - \nu,
\end{align}
\end{subequations}
with the quantities $g(\nu)$ and $h(\nu)$ given explicitly by Eqs.~(\ref{eq:gFactor}) and~(\ref{eq:hFactor}).

The system~(\ref{eq:reducedAMEs}) can be solved numerically to obtain $\rho(t)$ and thus characterize the temporal evolution of the adoption process. Let us further separate the fraction of adopters as $\rho(t) = \rho_0(t) + \rho_1(t)$, where $\rho_0$ and $\rho_1$ are the fractions of innovators and induced adopters (i.e. vulnerable and stable nodes), respectively. Now consider the identity,
\begin{equation}
\label{eq:suscepIdent}
1 - \rho = \sumk \Pk \summ \skm = r + (1 - r) \sumkc P(k) P(c) \summ \skm = r + \rho_s,
\end{equation} 
where $\rho_s(t)$ is the fraction of non-immune, susceptible nodes that can eventually adopt, either spontaneously or not. Since such susceptible nodes adopt spontaneously at a rate $\pr$, the rate equation for innovators is $\dot{\rho}_0 = \pr \rho_s$. Then, with Eq.~(\ref{eq:suscepIdent}) we obtain,
\begin{equation}
\label{eq:innovRateEq}
\rho_0(t) = \pr \int_0^t [1 - r - \rho(t)] dt,
\end{equation}
which can be calculated explicitly with the numerical solution of Eq.~(\ref{eq:reducedAMEs}).

\section{Waiting time of adoption}
\label{sec:tw}

\begin{figure}[!ht] \centering
\subfigure[]{
  \includegraphics[width=76mm]{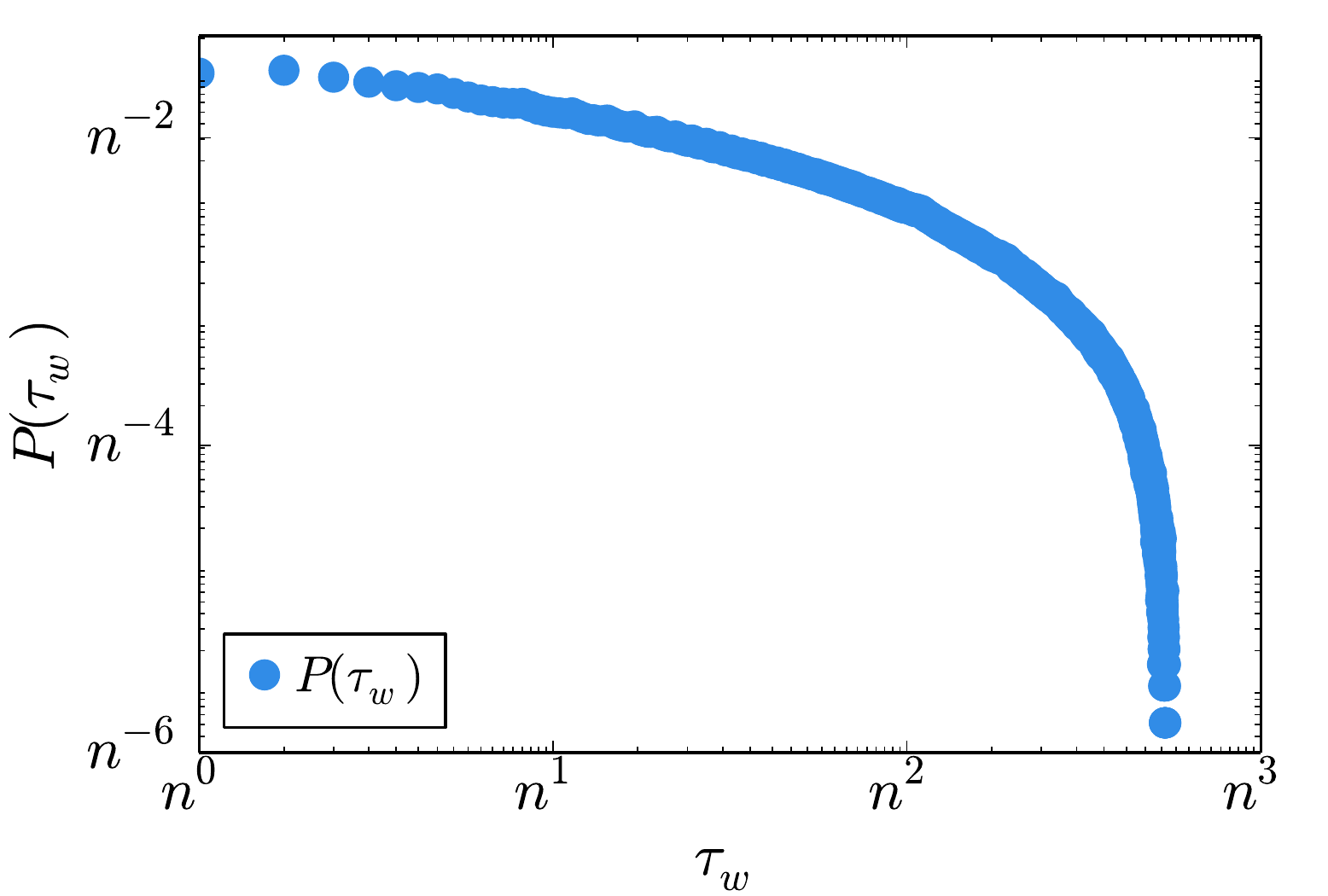}
}  
\subfigure[]{
  \includegraphics[width=76mm]{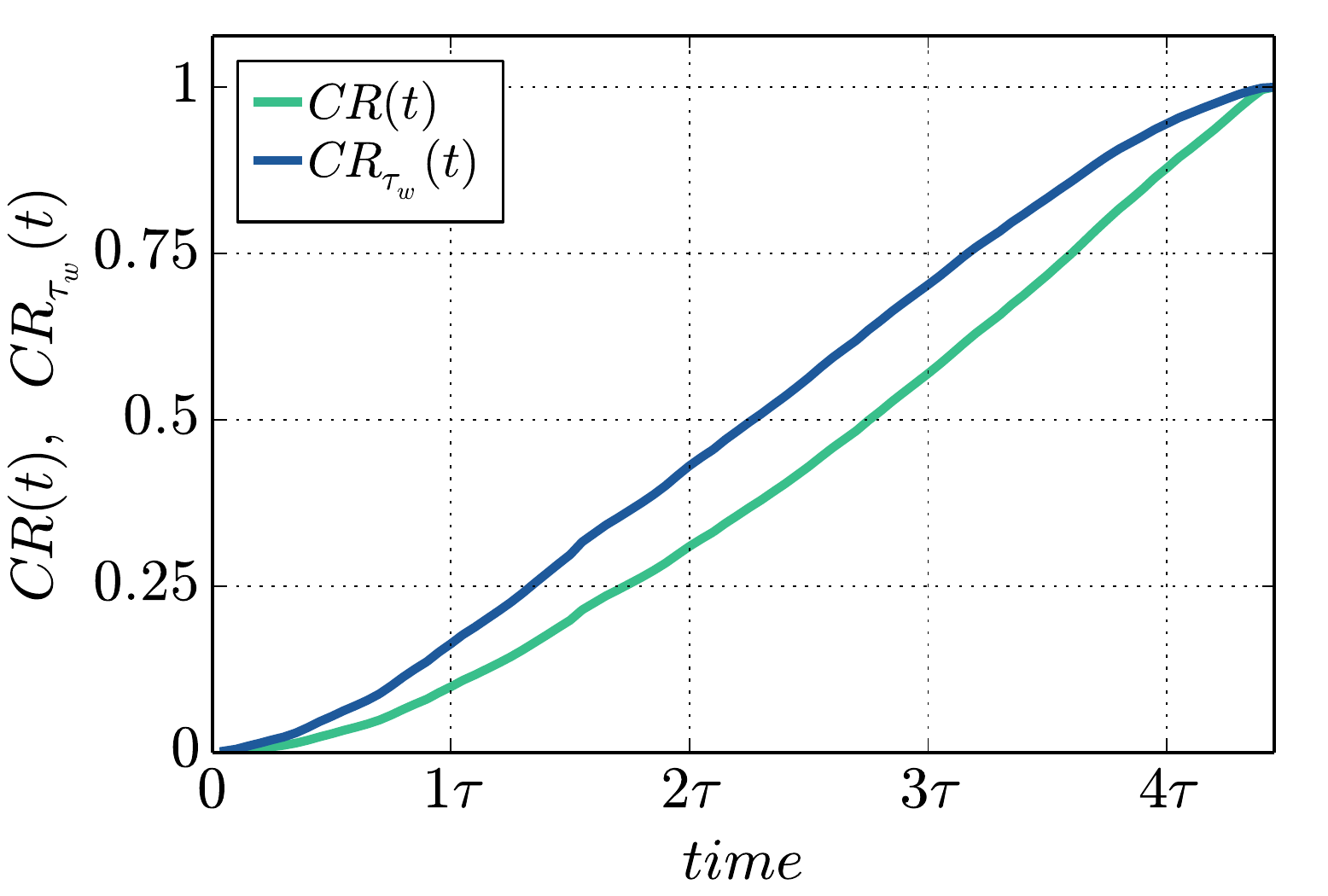}
}\\  
\caption{\small{\bf Waiting time distribution and its effect on the adoption process} {\bf (a)} Distribution $P(\tau_w)$ of times between the last adoption in the egocentric network of an individual and his/her own adoption. {\bf (b)} Cumulative adoption rates after waiting time removal [$CR(t)$ and $CR_{\tau_w}(t)$, respectively]. $n$ and $\tau$ are arbitrary constant values.}
\label{fig:WaitingT}
\end{figure}

Another reason behind the non-rapid evolution of the adoption process could be the time users wait after their personal adoption threshold is reached and before adopting the service. This lag in adoption can be due to individual characteristics, or come from the fact that social influence does not spread instantaneously (as commonly assumed in threshold models, including ours). However, the waiting time $\tau_w$ can be estimated by measuring the time difference between the last adoption in a user's egocentric network and the time of adoption. We define $\tau_w=0$ for innovators, but $\tau_w$ can take any positive value for vulnerable and stable adopters up to the length of the observation period.

Waiting times are broadly distributed for adopters (Fig.~S\ref{fig:WaitingT}a), meaning that many users adopt the service shortly after their personal threshold is reached, but a considerable fraction waits long before adopting the service. The heterogeneous nature of waiting times may be a key element behind the observed adoption dynamics. One way to figure out the effect of waiting times on the speed of cascade evolution is by removing them. We can extract waiting times from adoption times and thus calculate rescaled adoption times. The rescaled adoption time of a user is the last time when his/her fraction of adopting neighbours changed and the adoption threshold was (hypothetically) reached. After this procedure we can calculate a new adoption rate function by using rescaled adoption times and compare it to the original. From Fig.~S\ref{fig:WaitingT}b we can conclude that although adoption becomes faster, the rescaled adoption dynamics is still not rapid. On the contrary, it suggests that the rescaled adoption dynamics is still very slow and quite similar to the original. Consequently, waiting times cannot explain the observed dynamics of adoption.

Note that long waiting times can have a further effect on the measured dynamics. After the `real' threshold of a user is reached and he/she waits to adopt, some neighbours may adopt the product. Hence all observed measures are in this sense `effective': observed thresholds are larger or equal than real thresholds; the innovator rate is smaller or equal; the vulnerable and stable rates will be larger or equal; and waiting times will be shorter or equal than the real values. Consequently the process may be actually faster than that we observe in Fig.~S\ref{fig:WaitingT}b after removing effective waiting times. However, this bias becomes important only after the majority of individuals in the social network has adopted the service and the spontaneous emergence of adopting neighbours becomes more frequent. As the fraction of adopters in our dataset is always less than $6\%$ \cite{SkypeIPO}, we expect minor effects of this observational bias on measurements.

\section{Empirical and model cluster statistics}
\label{sec:clustStats}

\begin{figure}[!ht] \centering
\subfigure[]{
  \includegraphics[width=76mm]{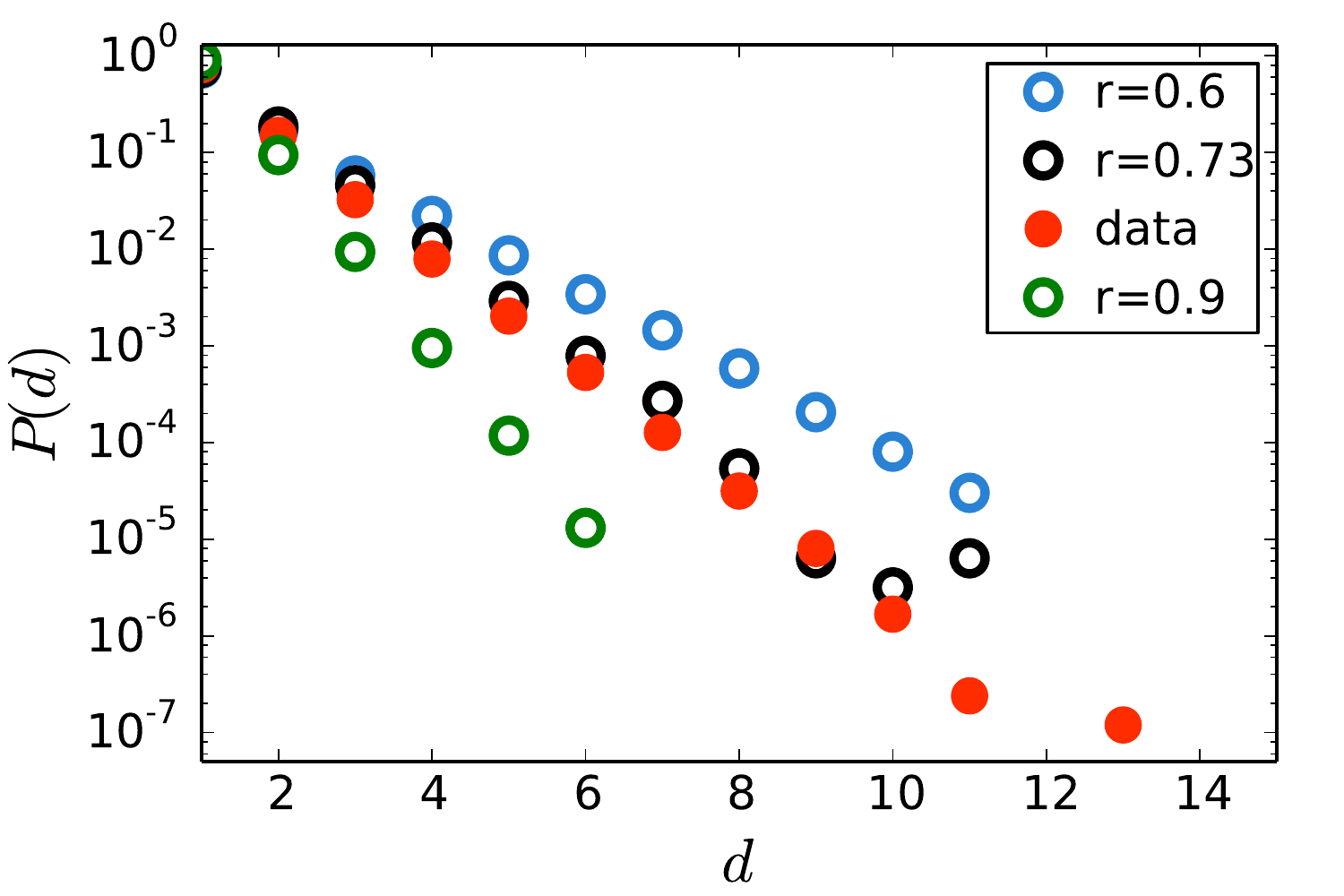}
}  
\subfigure[]{
  \includegraphics[width=76mm]{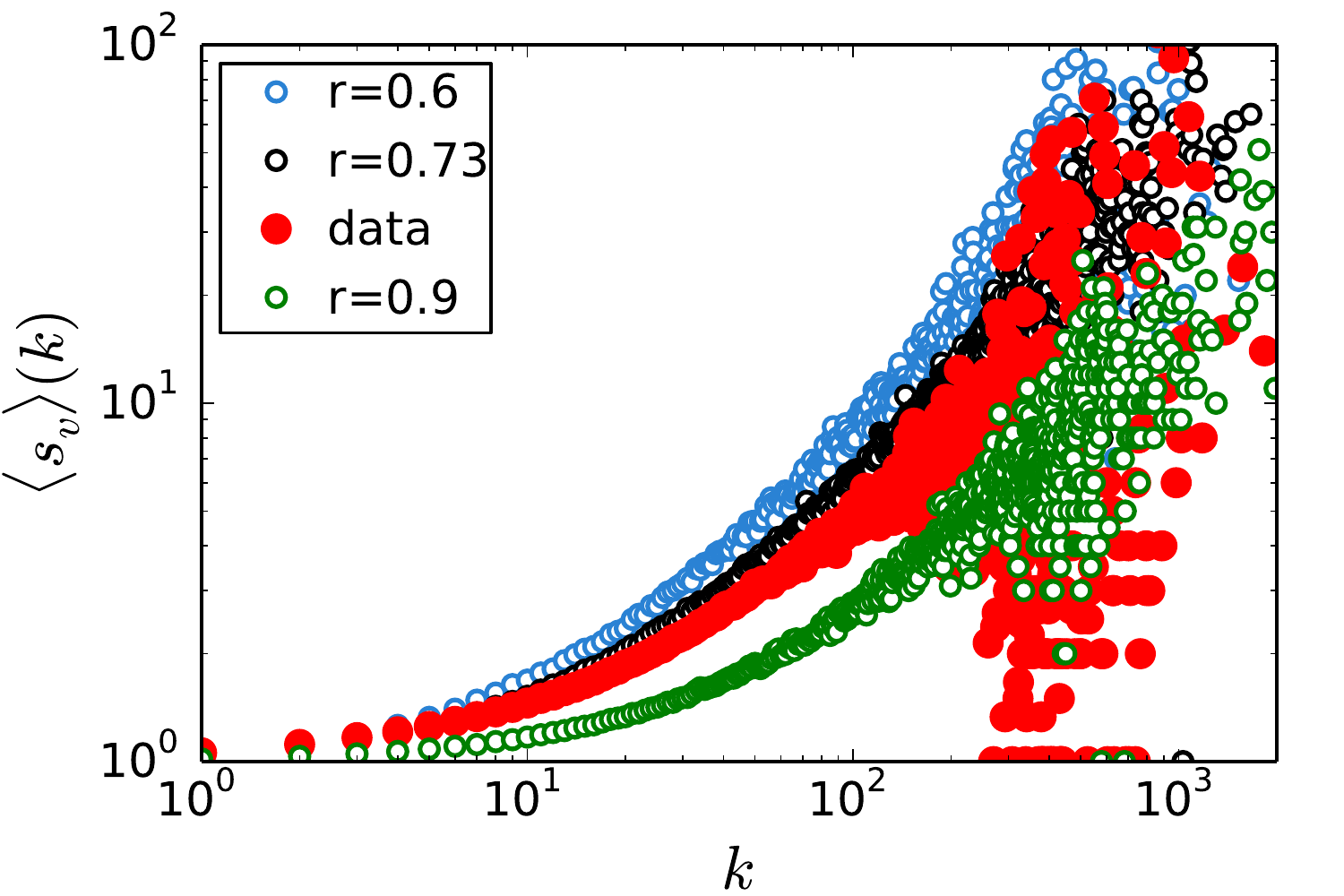}
}\\  
\caption{\small{\bf Empirical and model cluster statistics} {\bf (a)} Distribution $P(d)$ of the depth of induced vulnerable trees in the empirical and model systems. {\bf (b)} Correlation $\langle s_v \rangle (k)$ between the degree of innovators and the average size of vulnerable trees they induce. Empty symbols denote model calculations for $r=0.6$ (blue), $0.73$ (black), and $0.9$ (green), and full red symbols the empirical measurements. Model calculations correspond to networks of size $N=10^6$ and are averaged over $100$ independent realizations.}
\label{fig:ClusStat}
\end{figure}

As described in the main text, we perform extensive model calculations using empirically determined parameters to estimate the only unknown parameter, the fraction of immune nodes $r$. We match the relative size of the largest connected component of the real adoption network with its corresponding measure in the model at the end of the observation period, and estimate the fraction of immune nodes in the real system as $r=0.73$. To support our estimation we also measure the distribution $P(d)$ of the depth of induced vulnerable trees and the correlation $\langle s_v \rangle (k)$ between the degree of innovator nodes and the average size of induced vulnerable trees in the model, and match them with the equivalent empirical measures. To provide further support for the estimated $r$ value we show the dependence of these quantities of different $r$ values.

We measure $P(d)$ and $\langle s_v \rangle (k)$ for $r=0.6$ and $0.9$, as well as for the predicted value $r=0.73$ (Fig.~S\ref{fig:ClusStat}). It is clear that both quantities scale with $r$. For smaller $r$ more nodes are susceptible for adoption, allowing deeper and larger vulnerable trees, while for larger $r$ no large induced cluster can emerge as the system is forced into a quenched state. Moreover, measures for the estimated $r$ value fit the empirical data considerably well. This collapse is remarkable, since we neglect any higher-order structural and temporal correlations in the model (like assortative mixing, community structure, bursty adoption patterns, periodic activity fluctuations, etc.), which are present in the empirical system. Differences in the tails of the measures are due to finite-size effects since the modelled network is two orders of magnitude smaller than the empirical social structure. Note that although we can look for an $r$ fraction that produces a better fit between model and data in terms of $P(d)$ and $\langle s_v \rangle (k)$, the collapse in Fig.~S\ref{fig:ClusStat} demonstrates the quality of an independent procedure of estimating $r$ (i.e. by matching the relative size of components). Therefore, these results are intended for validation only and not as a method to estimate the correct value of $r$.

\section{Calculations for additional service}
\label{sec:addserv}

\begin{figure}[!ht]
\centering
\subfigure[]{
  \includegraphics[width=76mm]{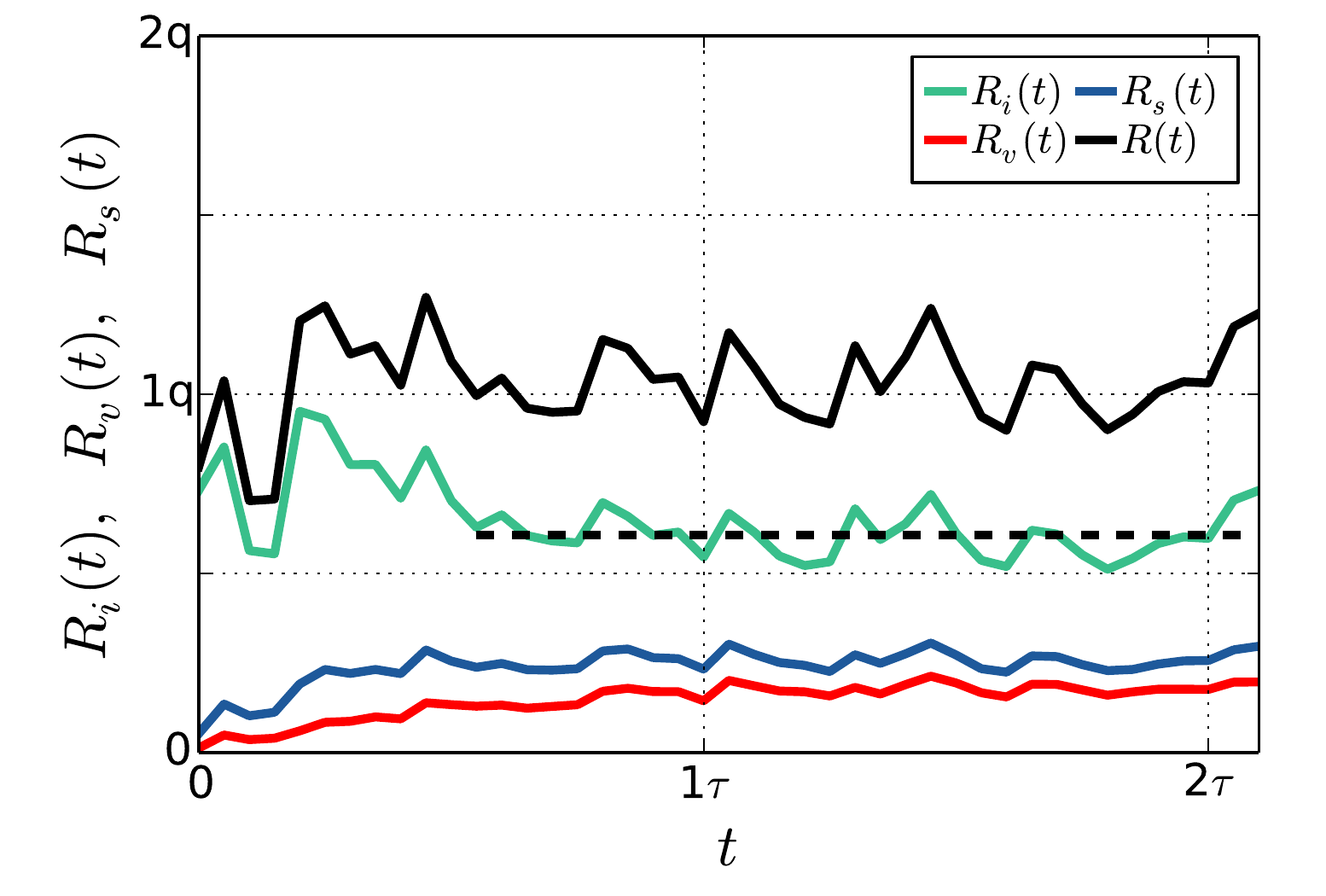}
}  
\subfigure[]{
  \includegraphics[width=76mm]{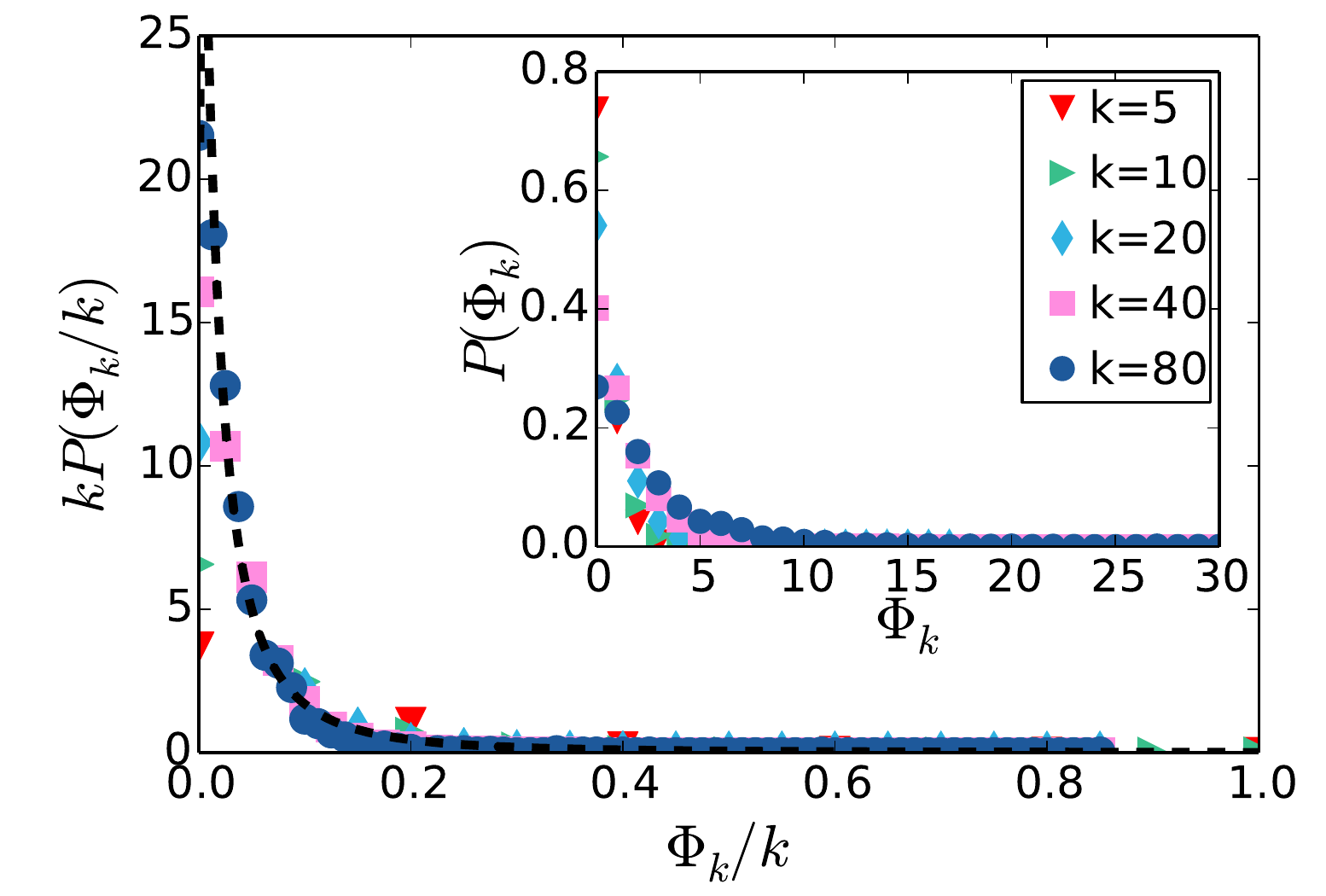}
}\\  
\caption{\small{\bf Adoption rates and threshold distributions for service ``subscription''}. {\bf (a)} Net adoption rate (black), as well as rates for innovator (green), vulnerable (red), and stable (blue) nodes as function of time. The dashed line is a fitted constant function to estimate the innovator adoption rate as $p_n=0.00012$. {\bf (b)} Distribution of integer thresholds $\Phi_k$ for several degree groups (inset). By using $P(\Phi_k,k)=k P (\Phi_k/k)$ these curves collapse to a master curve well approximated by a lognormal function (dashed line) with average $w=0.063$ and STD $0.153$ (for further details see Section \ref{sec:thrDistr}).}
\label{fig:srv4RateTh}
\end{figure}

\subsection{Empirical observations}

In order to support our empirical observations and modelling of the social spreading of Skype, we examine the adoption dynamics of an additional paid service called ``subscription'', introduced in April 2008 and with adoption data for over $42$ months until the end of the observation period. This service is only available for registered Skype users, and we can therefore use the accumulated static Skype network as background social structure. In order to investigate the adoption of this service we repeat all calculations described previously. First we measure the decoupled rate of innovator, vulnerable, and stable adopters (Fig.~S\ref{fig:srv4RateTh}a). We see that after a short initial period innovators adopt approximately with a constant rate, setting the model parameter to $p_n=0.00012$. Moreover, here innovators dominate social spreading since the rate of vulnerable and stable adoptions is relatively low.

We also measure the integer threshold distribution for different degree groups (Fig.~S\ref{fig:srv4RateTh}b, inset) just as described in Section \ref{sec:thrDistr}. These distributions scale together after normalization with the scaling relation $P(\Phi_k,k)=k P (\Phi_k/k)$ (Fig.~S\ref{fig:srv4RateTh}b, main panel) and are well approximated by a lognormal distribution [Eq.~(\ref{eq:thrLogn})] with parameters $\mu_T=-3.73$ and $\sigma_T=1.39$, as determined by the average threshold $w=0.063$ and STD $0.153$. Note that since the adoption dynamics of this service is dominated by innovators, the average threshold $w$ is smaller than in the case of the ``buy credit'' service. All parameters are summarized in Table \ref{table:parssrv4}. Since the background network is the same for both services, network parameters are those of Table.\ref{table:pars}.

\begin{figure}[!ht]
\centering
\subfigure[]{
  \includegraphics[width=76mm]{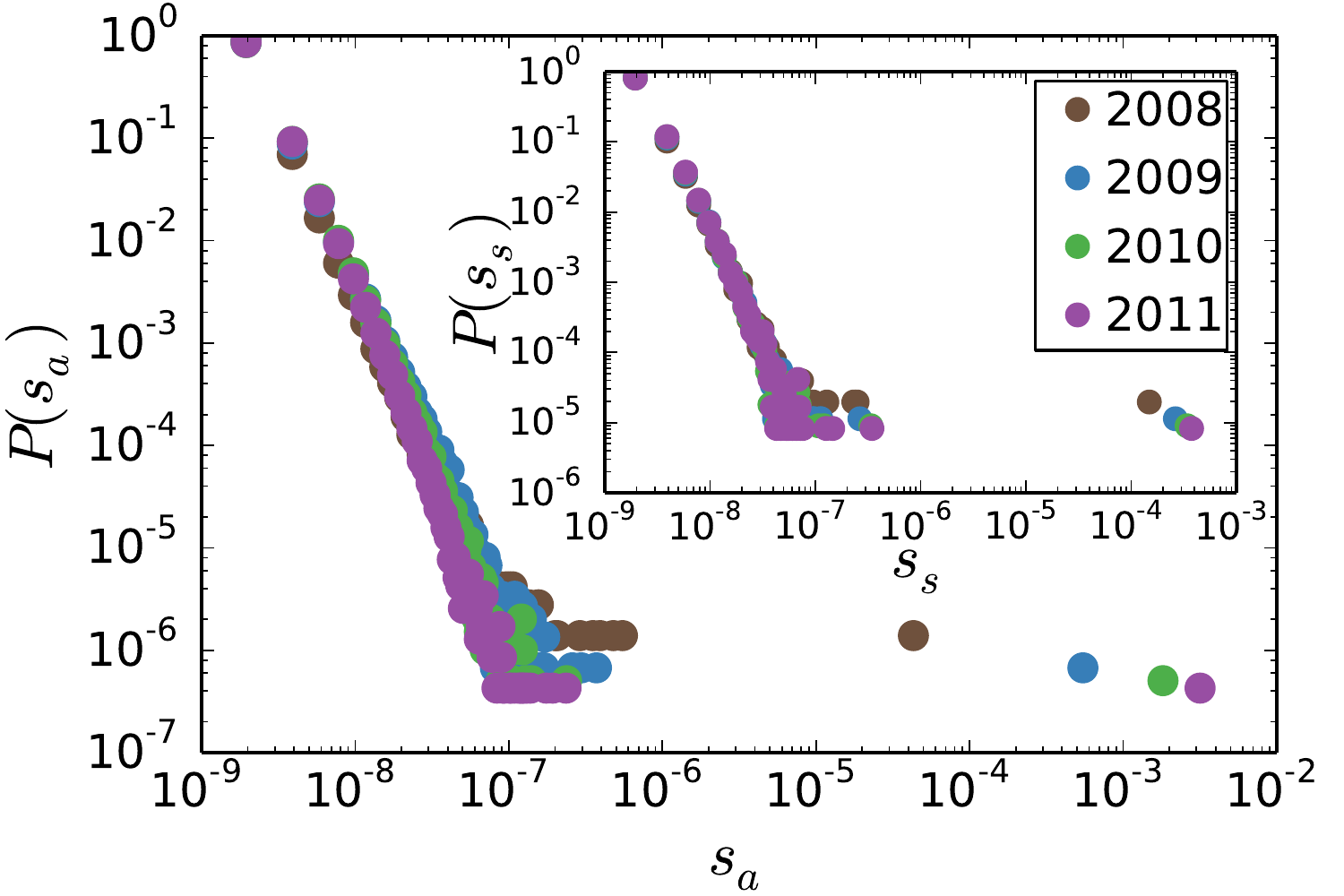}
}  
\subfigure[]{
  \includegraphics[width=76mm]{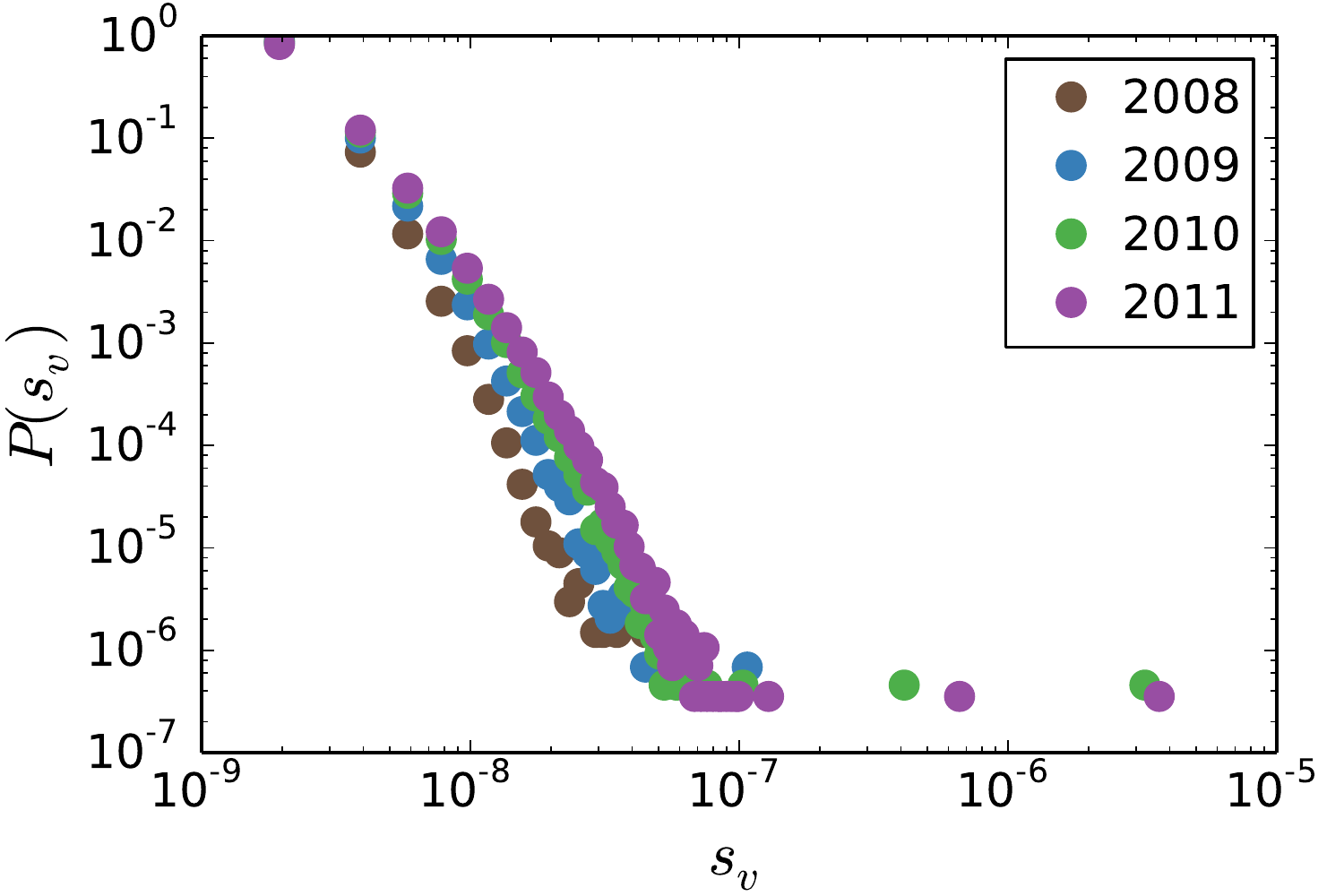}
}\\  
\caption{\small{\bf Empirical cluster statistics.} {\bf (a)} Relative connected-component size distribution $P(s)$ at different times for the empirical adoption network (main panel) and the stable adoption network (inset), with sizes $s_a$ and $s_s$, respectively. {\bf (b)} Relative connected-component size distribution $P (s_v)$ of the empirical innovator-induced vulnerable trees at different times.}
\label{fig:srv4adClust}
\end{figure}

Although the adoption process is dominated by innovators, a giant connected component evolves in the adoption network (Fig.~S\ref{fig:srv4RateTh}a, main panel). On the other hand, its relative size is considerable smaller than for the ``buy credit'' service. The stable adoption network is also dominated by a giant component, but its relative size is even smaller when compared to the adoption network (Fig.~S\ref{fig:srv4RateTh}a, inset). Moreover, the largest vulnerable trees are only two orders of magnitude smaller than the stable giant cluster (Fig.~S\ref{fig:srv4RateTh}b). For comparison, this difference is five order of magnitude for the ``buy credit'' service.

\begin{table}[h]
\begin{center}
\begin{tabular}{|c||c|c|c|c|}
\hline 
$p_{n}$ & $w$ & $STD(\phi)$ & $\mu_T$ & $\sigma_T$  \\ \hline \hline 
$0.00012$ & $0.063$ & $0.153$ & $-3.73$ & $1.39$ \\
\hline 
\end{tabular} 
\caption{\small Estimated empirical parameters for service ``subscription''.}
\label{table:parssrv4}
\end{center}
\end{table}

\subsection{Model and validation}

\begin{figure}[!ht]
\centering
\includegraphics[width=110mm]{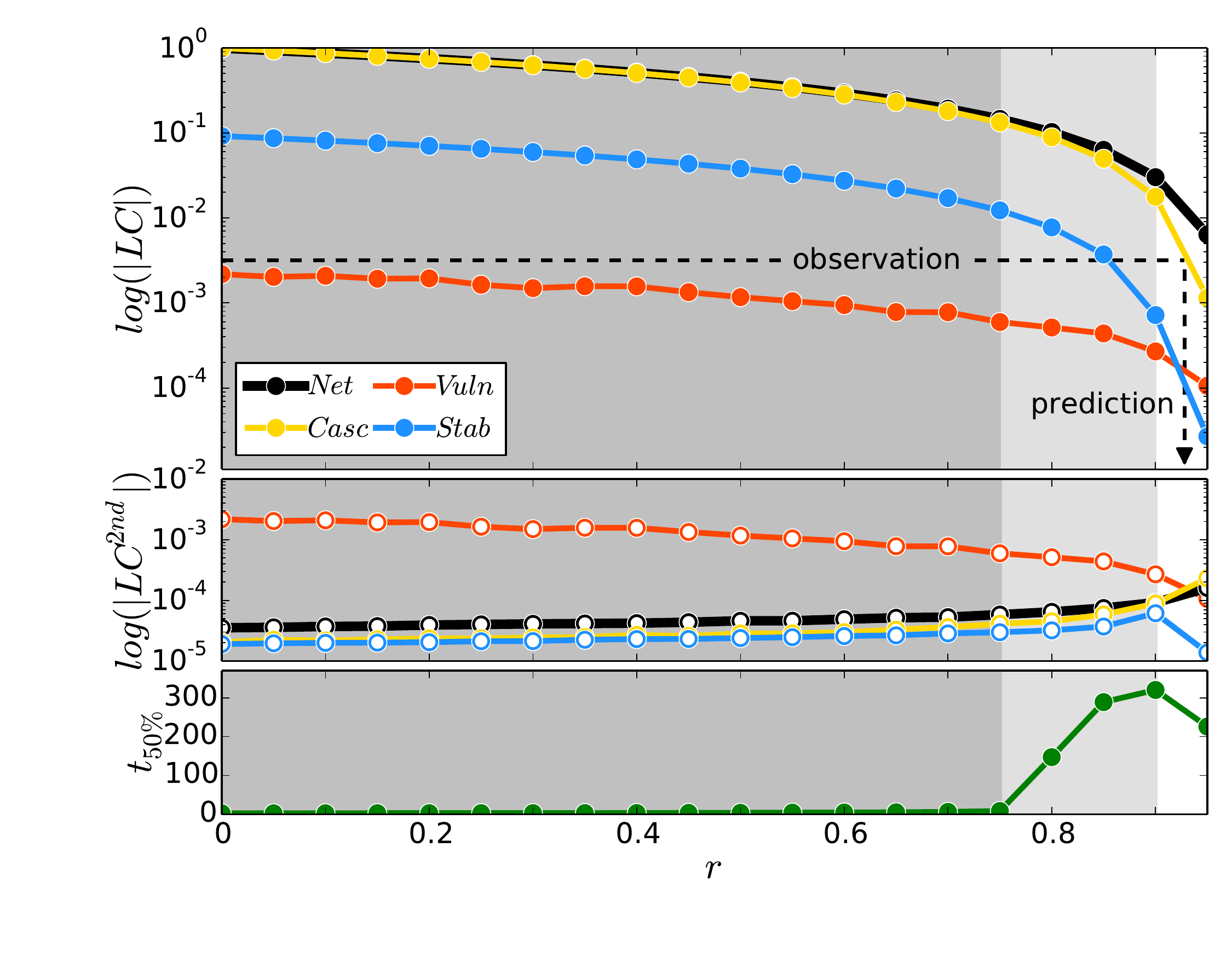}
\caption{\small{\bf Modeled adoption process of the service ``subscription''} Average size of the largest ($LC$) and 2nd largest ($LC^{2nd}$) components of the model network (`Net'), model adoption network (`Casc'), model stable network (`Stab'), and induced vulnerable trees (`Vuln') as a function of $r$. Dashed lines show the observed relative size of the real $LC$ of the adopter network in 2011 (Fig.~S\ref{fig:srv4adClust}, main panel) and the predicted $r$ value. The lower panel depicts the time $t_{50\%}$ when the adoption process has reached $50\%$ of the susceptible network as a function of $r$. We use $100$ realizations of configuration-model networks with size $N=10^5$ and lognormal degree distribution parametrized as described in Section \ref{sec:degDistrFit}. Model calculations correspond to the parameters of Table \ref{table:parssrv4} for $42$ iteration steps (matching the length of the observation period).}
\label{fig:srv4model}
\end{figure}

We repeat all model calculations with the parameters of the ``subscription'' service to see whether we can recover its adoption dynamics by using the dynamical threshold model introduced in the main text and in Section \ref{sec:thresModel}. We check the dependence on $r$ of the average size of the largest connected component of the network ($LC$) of susceptible nodes available for the adoption process, the adoption network, the stable adoption network, and of vulnerable trees (Fig.~S\ref{fig:srv4model}, upper panel). In addition we record the average size $LC^{2nd}$ of the second largest connected component (Fig.~S\ref{fig:srv4model}, middle panel). Finally we show the time when the adoption process has reached the $50\%$ of available susceptible nodes in the adoption network (Fig.~S\ref{fig:srv4model}, lower panel).

The $r$ dependence of the adoption process appears to be qualitatively similar to our earlier calculations on the ``buy credit'' service, but there are remarkable differences. Firstly, the crossover regime (depicted by the light grey area in Fig.~S\ref{fig:srv4model}) is shifted towards larger $r$ values due to the different threshold distribution and innovator adoption rate. Secondly, after matching the relative size of the largest connected component of the empirical adoption network (last point on the right-hand side of Fig.~S\ref{fig:srv4adClust}, main panel), the predicted $r=0.928$ is out of the crossover regime. At this point the background social network is still not fragmented (as evidenced by the black line in Fig.~S\ref{fig:srv4model}, which has not reached its maximum yet) and it allows for the emergence of large connected adoption clusters. It is very sparse, however, which explains: (a) the dominating innovator adoption rate observed empirically; (b) the reduced size of the giant component of the adoption and stable adoption networks; and (c) the relatively large innovator trees as compared to the stable adoption network components. We observe that the largest vulnerable trees are smaller than the largest stable clusters in the empirical data, while the opposite is true for the model. A possible explanation of this difference is the assumption in the model that the network is degree-uncorrelated. This is a necessary approximation in order to treat the model analytically, but it might not hold for the empirical network. All in all, this picture suggests that the ``subscription'' service is out of the rapid and even the crossover cascading regimes, and that its dynamics is mostly driven by independent innovators rather than social influence, on a network of which a large majority is not susceptible to innovation.

\end{document}